\documentclass[12pt]{article}
\newcommand{\be}{\begin{equation}}
\newcommand{\ee}{\end{equation}}
\newcommand{\bea}{\begin{eqnarray}}
\newcommand{\eea}{\end{eqnarray}}
\newcommand{\bean}{\begin{eqnarray*}}

\newcommand{\eean}{\end{eqnarray*}}
\font\upright=cmu10 scaled\magstep1
\newcommand{\ssf}{\sans}
\newcommand{\stroke}{\vrule height8pt width0.4pt depth-0.1pt}
\newcommand{\Z}{\hbox{\upright\rlap{\ssf Z}\kern 2.7pt {\ssf Z}}}

\newcommand{\C}{{\rlap{\rlap{C}\kern 3.8pt\stroke}\phantom{C}}}
\newcommand{\R}{\hbox{\upright\rlap{I}\kern 1.7pt R}}
\newcommand{\N}{\hbox{\upright\rlap{I}\kern 1.7pt N}}
\newcommand{\CP}{\C{\upright\rlap{I}\kern 1.5pt P}}
\newcommand{\PP}{\hbox{\upright\rlap{I}\kern 1.5pt P}}

\newcommand{\identity}{{\upright\rlap{1}\kern 2.0pt 1}}

\newcommand{\sbf}{\mbox{\boldmath $\sigma$}}

\newcommand{\HH}{\mbox{\hbox{\upright\rlap{I}\kern 1.7pt H}}}

\newcommand{\fr}{\frac}
\newcommand{\lm}{\lambda}
\newcommand{\ra}{\rightarrow}
\newcommand{\al}{\alpha}

\newcommand{\pr}{\partial}
\newcommand{\hs}{\hspace{5mm}}

\newcommand{\dg}{\dagger}

\newcommand{\ve}{\varepsilon}
\newcommand{\acc}{\\[3mm]}

\input{epsf}

\begin{document}
\begin{center}{\bf
Non-trivial Soliton Scattering in Planar Integrable Systems
}\\
\vglue 0.5cm {Theodora Ioannidou{\footnote{{\it Email}:
T.Ioannidou@ukc.ac.uk\\ To appear in: {\it Internation Journal of Modern Physics 
A}}}, } \\
\vglue 0.3cm {\it Institute of Mathematics,  University
of Kent, Canterbury CT2 7NF, UK }\\
\end{center}

\begin{abstract}

The behavior of solitons in integrable theories is strongly
constrained by the integrability of the theory, that is by the
existence of an infinite number of conserved quantities that these
theories are known to possess. As a result the soliton scattering
of such theories are expected to
 be trivial (with no change of direction, velocity or shape).
In this paper we present an extended review on soliton scattering
of two spatial dimensional integrable systems which have been
derived as dimensional reductions of the self-dual Yang-Mills-Higgs
 equations and whose scattering properties are highly non-trivial.
\end{abstract}

\renewcommand{\thefootnote}{\arabic{footnote}}
\setcounter{footnote}{0}

\section{ Introduction}

Solitons which scatter in a non-trivial way and which occur as
solutions of planar integrable systems are going to be presented.
Initially  {\it solitons} were introduced by mathematicians to
describe lumps of energy stable to perturbations which do not
change either their velocity or their shape when colliding with
each other. They have been observed experimentally both as waves
on shallow water and in laser pulses in fibre-optic cables, among
other places. Solitons also play an important role in various
models of subatomic particles (quantum field theories). In recent
literature all sorts of localized energy configurations have been
called solitons which travel without assuming stability of the
shape or velocity or a simple behavior in collision.

In one space dimension integrable systems occur when dispersion
effects are exactly balance by nonlinearities. However, in more
than one dimensions the definition of an integrable model  is
related with the existence of infinite number of conserved
quantities;  of Painlev\'e property;  of two linear equations
 (so-called the Lax pair) which compatibility condition give
the non-linear soliton equation;  of inverse-scattering transform;
even of multi-soliton solutions (see, for example, \cite{AC}).

An interesting problem is the investigation of  the scattering
properties of two or more solitons colliding. In some known models
with non-trivial topology the collision of two solitons is
inelastic (some radiation is emitted) and
 non-trivial, ie a head-on collision results in $90^0$ scattering
(see, for example, \cite{Z} and references therein).
On the other hand, the solitons  of integrable models
interact trivially in the sense that they pass through each other
with no lasting change in velocity, shape or direction.
Some examples in 2+1 dimensions are the Kadomtsev-Petviashvili
 \cite{KP}, the Konopelchenko-Rogers \cite{KR},
the Davey-Stewrtson \cite{DS} and the integrable chiral equation
\cite{W}. Until recently, non-trivial scattering of solitons occurs
mostly in non-integrable models which is far from simple. The
issue discussed here is whether such type of scattering can occur
in integrable models too. There are some limited examples of
integrable models where soliton dynamics is non-trivial. In 1+1
dimensions there are many models that non-trivial soliton-like
solutions (cf. Ref. \cite{1d}) exists like the boomeron solutions
\cite{Bo} which are solitons with time-dependent velocities. In
2+1 dimensions there are the dromion solutions \cite{dr} of the
Davey-Stewartson equation which decay exponentially in both
spatial coordinates and interact in a non-trivial matter
\cite{drs} and the soliton solutions of the
Kadomtsev-Petciashvilli equations whose scattering properties are
highly non-trivial \cite{KPs}.

Recently in  \cite{WI}-\cite{I2} families of soliton solutions
have been constructed for planar integrable systems (using
analytical methods) with different types of scattering behavior.
This happens since the solitons of the systems
 have internal degrees of  freedom that although they determine their space
orientation, they do not change the energy density and they are
important in understanding the system evolution. Therefore,
solitons can interact either trivially or non-trivially depending
on the orientation of these internal parameters and on the values
of the impact parameters defined as the distance of closest of
approach between their centers in the absence of interaction.

In what follows we will review the approach based on the
 {\it Riemann-Hiblert problem with zeros}, first presented in
\cite{WI}-\cite{I2}, in order to construct soliton solutions with
non-trivial scattering for two spatial dimensional integrable
models defined in  anti-de Sitter and Minkowski space,
respectively. Both models are equivalent  to  the self-dual
Yang-Mills-Higgs equations reduced from 2+2 to 2+1 dimensions
while the soliton solutions are related to hyperbolic and
Euclidean Bogomolny-Prasad-Sommerfield (BPS) monopoles.

\section{Self-dual Yang-Mills-Higgs Equations}

Static BPS monopoles are solutions of  non-linear elliptic partial
 differential equations on some three-dimensional Riemannian
manifold.
 Most work on monopoles has dealt with the case when this
manifold is the Euclidean space $\R^3$ since then the equations
are integrable and sophisticated geometrical techniques like
twistor theory can be applied. Note that the introduction of
time-dependence destroys the integrability. In addition, the monopole
equations on hyperbolic space $\HH^3$ are also integrable
\cite{At} and often hyperbolic monopoles turn out to be easier to
study than the Euclidean ones as first studied by Atiyah and
explicitly shown in \cite{IS}.
 In fact  as it has been rigorously
established in \cite{JN}, in the limit as the curvature of
hyperbolic space tends to zero the Euclidean monopoles are
recovered.

Let us first  consider  an integrable system \cite{W1} which is related to
hyperbolic monopoles and which is obtained from replacing the positive definite
 hyperbolic space $\HH^3$ by a Lorentzian version,  the so-called
 anti-de Sitter space.
Thus, the self-dual  Yang-Mills-Higgs equations are of the form
\be
 D_i \Phi=\fr{1}{2\sqrt{|g|}} g_{ij} \,\varepsilon^{jkl}
F_{kl}
 \label{Bog}
  \ee
   where the Yang-Mills-Higgs fields take values on a
three-dimensional Riemannian manifold (${\cal M}$) with gauge
group $SU(2)$.
 In particular, $A_k$ (for $k=0,1,2$) is the
$su(2)$-valued gauge potential,
  $F_{ij}=\pr_iA_j-\pr_jA_i+[A_i,A_j]$ is the  field strength and
$\Phi=\Phi(x^\mu)$ is the $su(2)$-valued Higgs field; while
 $x^\mu=(x^0,x^1,x^2)$ represent the local coordinates on ${\cal M}$.
The action of the covariant derivative $D_i=\pr_i+A_i$ on $\Phi$ is:
 $D_i\Phi=\pr_i\Phi+[A_i,\Phi]$.
Equation (\ref{Bog}) for constant curvature is integrable in the
sense that a Lax pair exists.

Note that  the solutions of (\ref{Bog}) which can be described in
terms of holomorphic vector bundles or in terms of rational
functions
 correspond to Euclidean  or hyperbolic BPS monopoles when
$({\cal M},g)$ is the Euclidean $\R^3$ or  the hyperbolic $\HH^3$
space, respectively. Also, both of the models presented here can
be obtained from (\ref{Bog}) as dimensional reductions of the
four-dimensional  self-dual Yang-Mills-Higgs equations for appropriate
gauge choices \cite{AC}.

\section{The Anti-de Sitter Model}

Currently a great deal of attention has been focused on anti-de
Sitter spaces since they arise naturally in black holes and
$p$-branes. For the case of Yang-Mills theory with ${\cal N}=4$
supersymmetries and a large number of colors it has been
conjectured that gauge strings are the same as the fundamental
strings
 but moving in a particular curved space: the product of
five-dimensional anti-de Sitter space and a five sphere
\cite{Mal}. Then, using Poincar\'e coordinates the anti-de Sitter
solutions play the role of classical  sources for the boundary
field correlators, as shown in \cite{Wit}; while extensions of the
corresponding statements can be applied to gravity theories, like
 black holes which arise in anti-de Sitter backgrounds.

By definition the (2+1)-dimensional anti-de Sitter space is the universal
covering space of the hyperboloid ${\cal H}$ defined by the equation
\be
U^2+V^2-X^2-Y^2=1
\ee
with metric
\be
ds^2=-dU^2-dV^2+dX^2+dY^2.
\ee
By parametrizing the hyperboloid ${\cal H}$ by
\bea
U&=&\sec\rho\cos\theta\nonumber\\
V&=&\sec\rho\sin\theta\nonumber\\
X&=&\tan\rho\cos\phi\nonumber\\
Y&=&\tan\rho\sin\phi
\eea
for $\rho \in [0,\pi/2)$, the corresponding metric simplifies to
\be
ds^2=\sec^2\rho \left(-d\theta^2+d\rho^2+\sin^2\rho\, d\phi^2\right).
\ee
The spacetime contains closed timelike curves due to the periodicity
of $\theta$ \cite{HE}.
In fact $(\rho,\phi)$ correspond to polar coordinates
and $\theta \in R$ being the time; while
 anti-de Sitter space (as a manifold) is the product of an
 open spatial disc with $\theta$ and  curvature equal
to minus six.
Null spacelike infinity ${\cal I}$ consists of the timelike
cylinder $\rho=\pi/2$ and this surface is never reached by
timelike geodesics.

\begin{figure}
\vskip 0.5cm
\centerline{
\put(120,95){$\theta$}
\put(100,50){$r=0$}
%\put(5,80){$r=\pi/2$}
\put(60,60){$t$}
\put(110,3){$\rho$}
\put(6,100){$(\rho,\phi,\theta)$}
\put(14,58){$(r,x,t)$}
\epsfxsize=4cm\epsfysize=4cm\epsffile{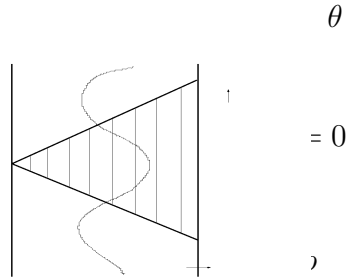}}
\vskip 0.5cm
\caption{The Penrose diagram of anti-de Sitter space.
The boundary of anti-de Sitter is the boundary of the cylinder.}
\label{fig-ads}
\end{figure}

If the Poincar\'e coordinates $(r,x,t)$ for $r>0$ are defined as
\bea
r&=&\fr{1}{U+X}\nonumber\\
x&=&\fr{Y}{U+X}\nonumber\\
t&=&\fr{-V}{U+X} \label{twis} \eea the metric simplifies to the
following form \be ds^2=r^{-2}(-dt^2+dr^2+dx^2). \label{Pmet} \ee
Note that the Poincar\'e coordinates cover a small part of anti-de
Sitter space which correspond to half of the hyperboloid ${\cal
H}$ for $U+X>0$ as shown in Figure \ref{fig-ads}. The surface $r=0$ is
part of infinity ${\cal I}$.

Consider the set of linear equations
\bea
\left[rD_r-2(\lm-u)D_u-\Phi\right]\Psi&=&0\nonumber\\
\left[2D_v+\fr{\lm-u}{r}D_r-\fr{\lm-u}{r^2}\Phi\right]\Psi&=&0.
\label{Lax} \eea
 Here  $\lm \in \C$ and $(r,u,v)$ are the
Poincar\'e coordinates (where $u=x+t$ and $v=x-t$) while the gauge
fields $(\Phi,A_r,A_u,A_v)$ are $2\times 2$ trace-free matrices
depending only on $(r,u,v)$ and $\Psi(\lm,r,u,t)$ is a unimodular
$2\times 2$ matrix function satisfying the reality condition
$\Psi(\lm)\Psi (\bar{\lm})^\dagger =I$ (where $\dagger$ denotes
the complex conjugate transpose). The system (\ref{Lax}) is
overdetermined and in order for a solution $\Psi$ to exist the
following integrability conditions need to be satisfied \bea
D_u \Phi&=&r F_{ur}\nonumber\\
D_v \Phi&=&-r F_{vr}\nonumber\\
D_r \Phi&=&-2r F_{uv} \label{sys} \eea which correspond to the
self-dual Yang-Mills-Higgs equations (\ref{Bog}) defined in the
2+1 anti-de Sitter space. The gauge and Higgs fields in terms of
the function $\Psi$ can be obtained from the Lax pair (\ref{Lax})
due to the boundary conditions. Note that, as  $\lm \ra \infty$
the function $\Psi$ goes to the identity matrix and the system
(\ref{Lax}) implies that \be A_u=0,\hs\hs A_r=\fr{1}{r}\,\Phi.
\label{G1} \ee On the other hand, for $\lm=0$ and using (\ref{G1})
the rest of the gauge fields are defined as \bea
\Phi&=&-\fr{r}{2}\,J_rJ^{-1}-u\,J_uJ^{-1}\nonumber\\
A_v&=&\fr{u}{2r}\,J_rJ^{-1}-J_vJ^{-1}
\label{G2}
\eea
where $J(r,u,v) \doteq \Psi(\lm=0,r,u,v)$.
As a result the first equation of the system (\ref{sys}) is
automatically satisfied (due to the specific gauge choice).

Recently, Ward \cite{W1} has shown that holomorphic vector
 bundles $V$ over $Q$
determine multi-soliton solutions of (\ref{sys}) in anti-de Sitter space
via the usual Penrose transform.
This way a five-parameter family of soliton solutions can be obtained,
in a similar way as for flat spacetime \cite{W2}.
Later, more solutions of equations (\ref{sys}) were obtained
by Zhou \cite{Zhou1,Zhou2} using Darboux transformations with constant
 and variable spectral parameter.
In what follows, we use the Riemann problem with zeros to
construct families of soliton solutions and observe the occurrence
of different types of scattering behavior. More precisely, we
present families of multi-soliton solutions with trivial and
non-trivial scattering.

\subsection{Baby Monopoles}

In this section, we illustrate the construction of time-dependent
solutions related to hyperbolic monopoles. In particular, families
of soliton solutions of the self-dual Yang-Mills-Higgs equations
defined in the (2+1)-dimensional anti-de Sitter space are
constructed and their dynamics is studied in some detail.

The integrable nature of (\ref{Bog}) means that there is a variety
of methods for constructing solutions. Here, we indicate a general
method for constructing soliton solutions of (\ref{sys}) which is
a variant of that presented in Ref. \cite{W2}. Using the standard
method of Riemann problem with zeros, in order to construct the
multi-soliton solutions, we assume that the function $\Psi$ of the
system  (\ref{Lax}) has
the simple form in $\lm$ \be \Psi=I+\sum_{k=1}^n
\fr{M_k}{\lm-\mu_k} \label{Psi} \ee
 where $M_k$ are $2\times2$
matrices independent of $\lm$ and $n$ corresponds to the soliton
number. The components of the matrix $M_k$ are given in terms of a
rational function $f_k(\omega_k)=a_k \,\omega_k+c_k$ of the
complex variable: $\omega_k=v-r^2\,(\mu_k-u)^{-1}$. Here $a_k$,
$c_k$ and $\mu_k$ are complex constants which determine the size,
position and velocity  of the $k$-th solitons. {\it Remark}: The
rational dependence of the solutions $\Psi$ follows (directly)
when the inverse spectral theory is considered. In fact in
\cite{FI} (for flat spacetime), it was shown from the Cauchy
problem that the spectral data is a function of a parameter
similar to $\omega_k$.

By considering the unitarity condition of the eigenfunction $\Psi$
and keeping in mind that the gauge fields are independent of the
spectral parameter $\lm$ it can be shown that the matrix $M_k$ has
the form \be M_k=\sum_{l=1}^n (\Gamma^{-1})^{kl} \bar{m}_a^l m_b^k
\label{M} \ee with $\Gamma^{-1}$ the inverse of \be
\Gamma^{kl}=\sum_{a=1}^2 (\bar{\mu}_k-\mu_l)^{-1}\bar{m}_a^km_a^l
\label{G} \ee and $m^k_a$ holomorphic functions of $\omega_k$ of
the form $m^k_a=(m_1^k,m_2^k)=(1,f_k)$. The Yang-Mills-Higgs
fields $(\Phi,A_r,A_v,A_u)$ can then be read off from
(\ref{G1}-\ref{G2}) and they automatically satisfy (\ref{sys})
while the corresponding solitons are spatially localized since
$\Phi \ra 0$ at spatial infinity.

\begin{figure}
\hskip .4cm
\hskip 2.6cm
\epsfxsize=8cm\epsfysize=8cm\epsffile{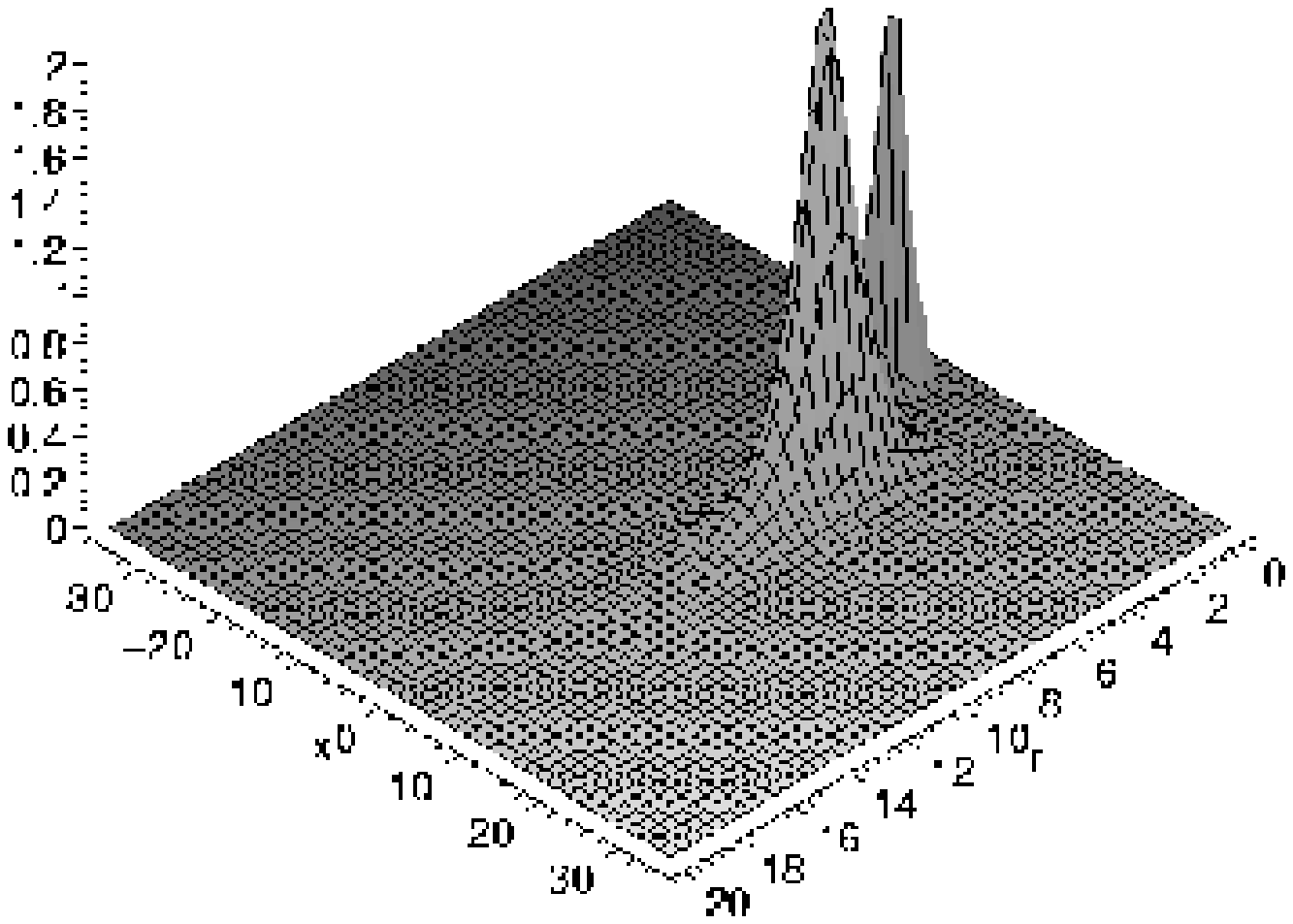}
\put(-280,160){$t=0$}
\par
\vskip -0.25cm
\hskip 4cm
\epsfxsize=8cm\epsfysize=8cm\epsffile{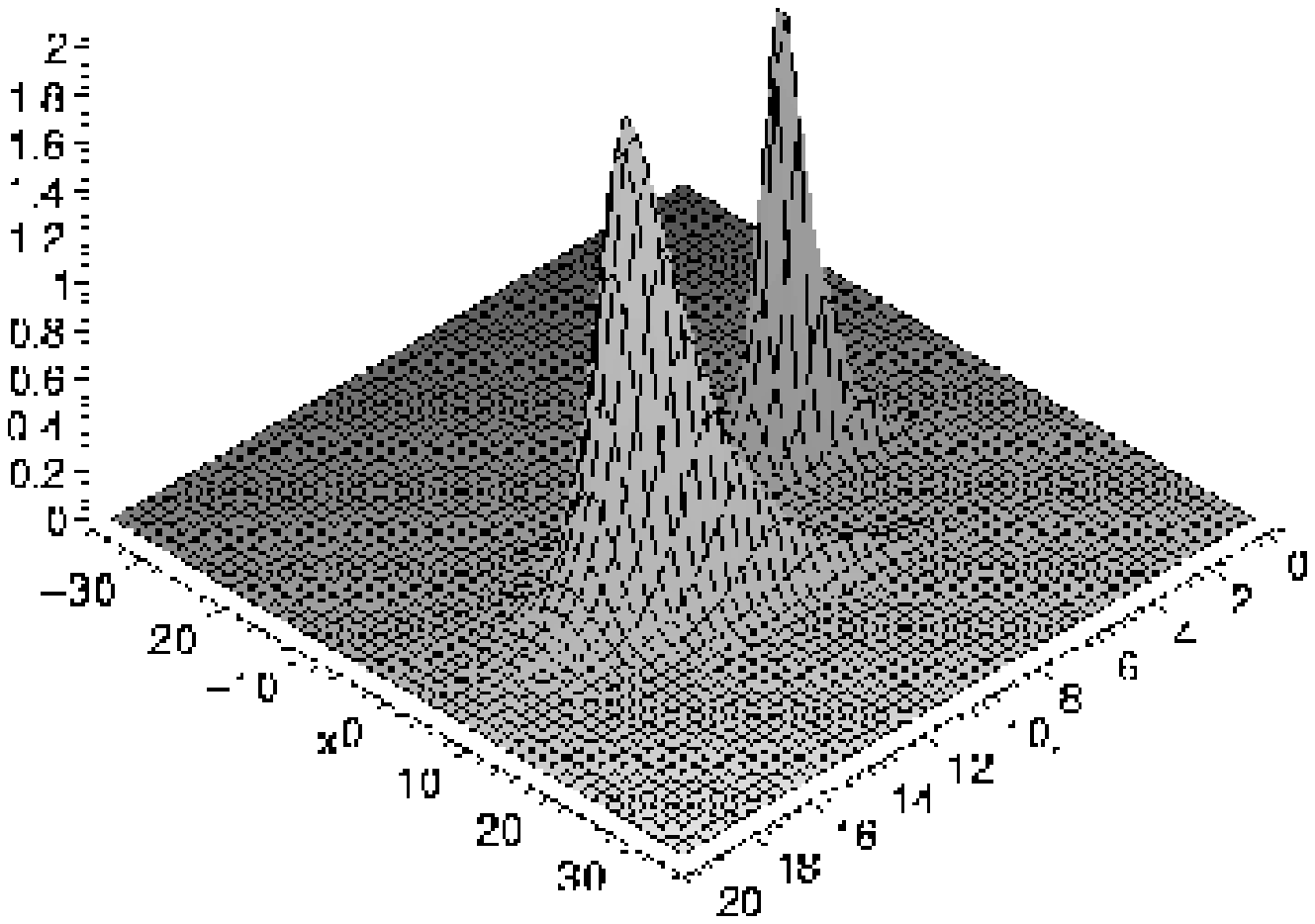}
\put(-260,160){$t=8$}
\vskip -0cm
\caption{A two-soliton configuration with trivial scattering at
different times.}
\label{fig-triv}
\end{figure}

By way of example, let us look at the special case where $\mu_1
=1+i$, $\mu_2=2i$, $a_1=2$, $a_2=1$, $c_1=5$ and $c_2=-10$. Figure
\ref{fig-triv} represents snapshots of the positive definite gauge
quantity ($-\mbox{tr} \Phi^2$) at different times. The
corresponding solution consists of two solitons which travel
towards infinity ($r=0$) and bounce back while their sizes change
as they move.

Note that the Riemann problem with zeros method was first
introduced by Zakharov and his collaborators \cite{Zak} in his
pioneer work of applying spectral theory to generate soliton
solutions of integrable equations. Later this approach has been
applied to obtain the monopole solutions of the four-dimensional
self-dual Yang-Mills-Higgs equations \cite{For}. However, as it
can been observed from (\ref{Psi}),  the Riemann problem with
zeros method assumes that the parameters $\mu_k$  are distinct and
also $\bar{\mu}_k \ne \mu_l$ for all $(k,l)$. In what follows
examples are given of two generalizations of these constructions:
one involving higher-order poles in $\mu_k$ and the other where
$\bar{\mu}_k = \mu_l$.\\

$\diamondsuit$\, {\it High-order poles in  $\mu_k$}\\

Firstly, let us look at an example in which the function $\Psi$
 has a double  pole in $\lm$ and no others.
In this case,  $\Psi$ has the form \be \Psi=I+\sum_{k=1}^2
\fr{R_k}{(\lm-\mu)^k} \ee where $R_k$ are $2\times 2$ matrices
independent of $\lm$. Then $\Psi$ corresponds to a solution of
(\ref{Lax}) if and only if it factorizes as \cite{I2}
 \be
\Psi(\lm)\!=\!\left(1-\fr{\bar{\mu}-\mu}{(\lm-\mu)} \fr{q^\dagger
 \otimes q}{|q|^2}\right) \!\left(1\!-\!\fr{\bar{\mu}-\mu}{(\lm-\mu)}
\fr{p^\dagger \otimes p}{|p|^2}\right) \label{Psi-sca} \ee
 for
some two vectors $q$ and $p$.
 One way to derive the structure of
these vectors is to take the formula (\ref{Psi}) for $n=2$, set
$\mu_1=\mu+\varepsilon$, $\mu_2=\mu-\varepsilon$ and
$f_1(\omega_1)= f(\omega_1)+\varepsilon h(\omega_1)$,
$f_2(\omega_2)=f(\omega_2)-\varepsilon h(\omega_2)$,  with $f$ and
$h$ being rational function of one variable. In the limit
$\varepsilon \ra 0$ the two vectors $q$ and $p$ can be obtained
and are of the form:
 \bea q&=&(1+|f|^2)(1,f)+(\bar{\mu}-\mu)\!
\left(\fr{r^2\,f'} {(\mu-u)^2}+h\!\right)
(\bar{f},-1),\nonumber\\
p&=&(1,f) \label{pq} \eea
 where $f$ and $h$ are rational functions of
 $\omega=v-r^2(\mu-u)^{-1}$ while $f'$ denotes the derivative of
 $f$ with respect to its argument.
 In this case, the solution
depends on the parameter $\mu$ and on the two arbitrary functions
$f$ and $h$.
  Note that the
constraint $f_2(\omega_2)-f_1(\omega_1)\ra 0$ as $\varepsilon \ra
0$ has to be imposed in order for the resulting solution $\Psi$ to
be smooth for  all $(r,u,v)$ which is the case here.

In order to illustrate the above family of solutions two
simple cases are going to be examined by giving specific values to the
parameters $\mu$, $f(\omega)$ and $h(\omega)$.

\begin{itemize}
\item
Let us study the simple case where $\mu=i$, $f(\omega)=\omega$ and
$h(\omega)=0$. Then, the quantity $-\mbox{tr} \Phi^2$ simplifies
to \bea -\mbox{tr}\Phi^2=32r^2
\fr{\left((r^2+x^2\!-\!t^2\!+\!1)^2+4t^2\right)
\left((r^2+x^2\!-\!t^2\!-\!1)^2+4x^2\right)}
{\left[\left((r^2+x^2-t^2)^2+1+2t^2+2x^2\right)^2+4r^4 \right]^2}
\eea which is time reversible. The corresponding time-dependent
solution is a travelling ring-like soliton configuration which for
negative $t$, goes towards spatial infinity $(r=0)$; approaches it
at $t=0$  and then bounces back at positive $t$ while the soliton
size deforms, as shown in  Figure \ref{fig-station}. Ring
structures occur in the soliton scattering of many non-integrable
planar systems and are approximations of multi-solitons
\cite{Zakp}.

\begin{figure}
\hskip .4cm
\hskip 2.6cm
\epsfxsize=8cm\epsfysize=8cm\epsffile{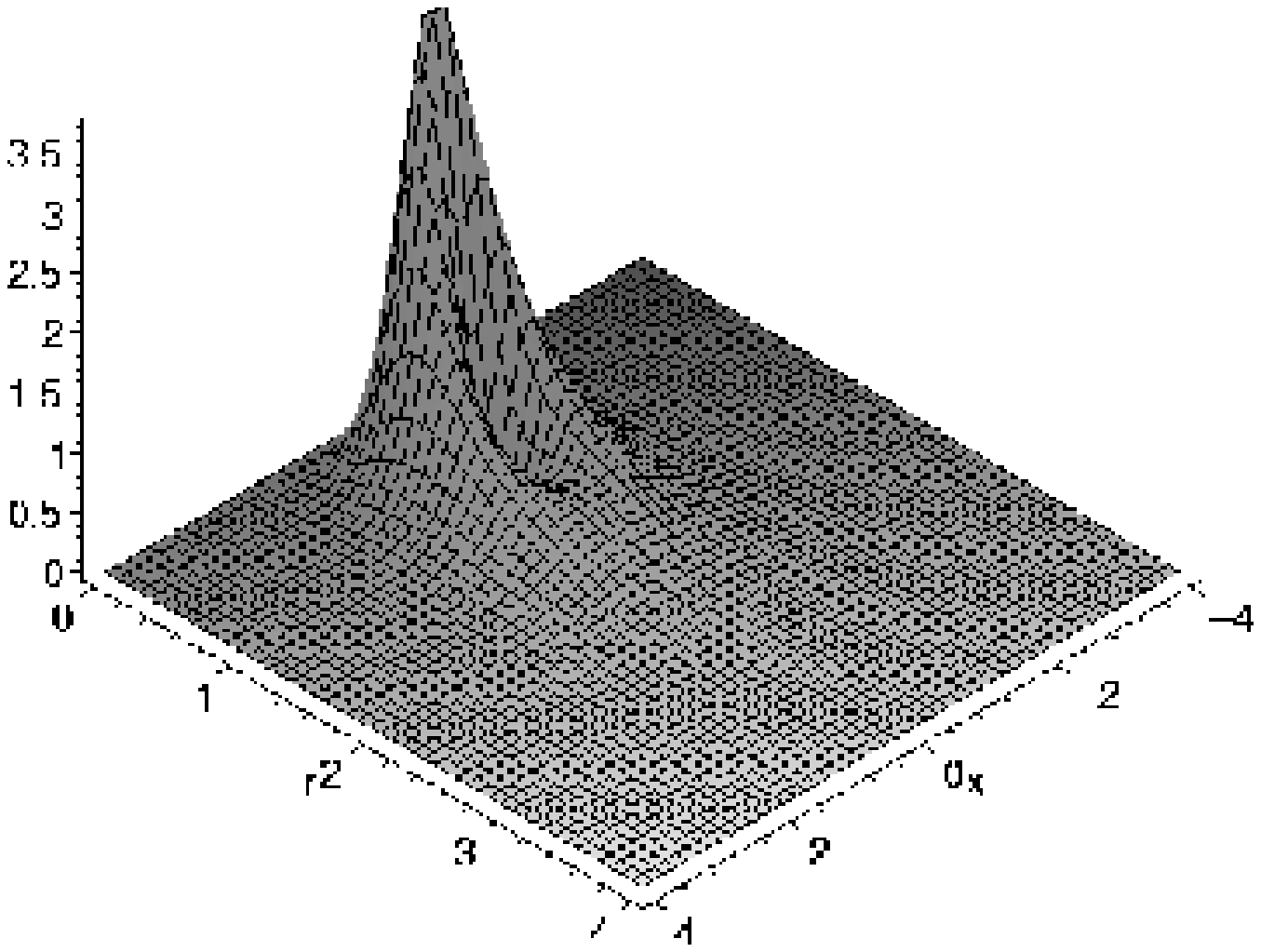}
\put(-280,160){$t=0$}
\par
\vskip -0.25cm
\hskip 4cm
\epsfxsize=8cm\epsfysize=8cm\epsffile{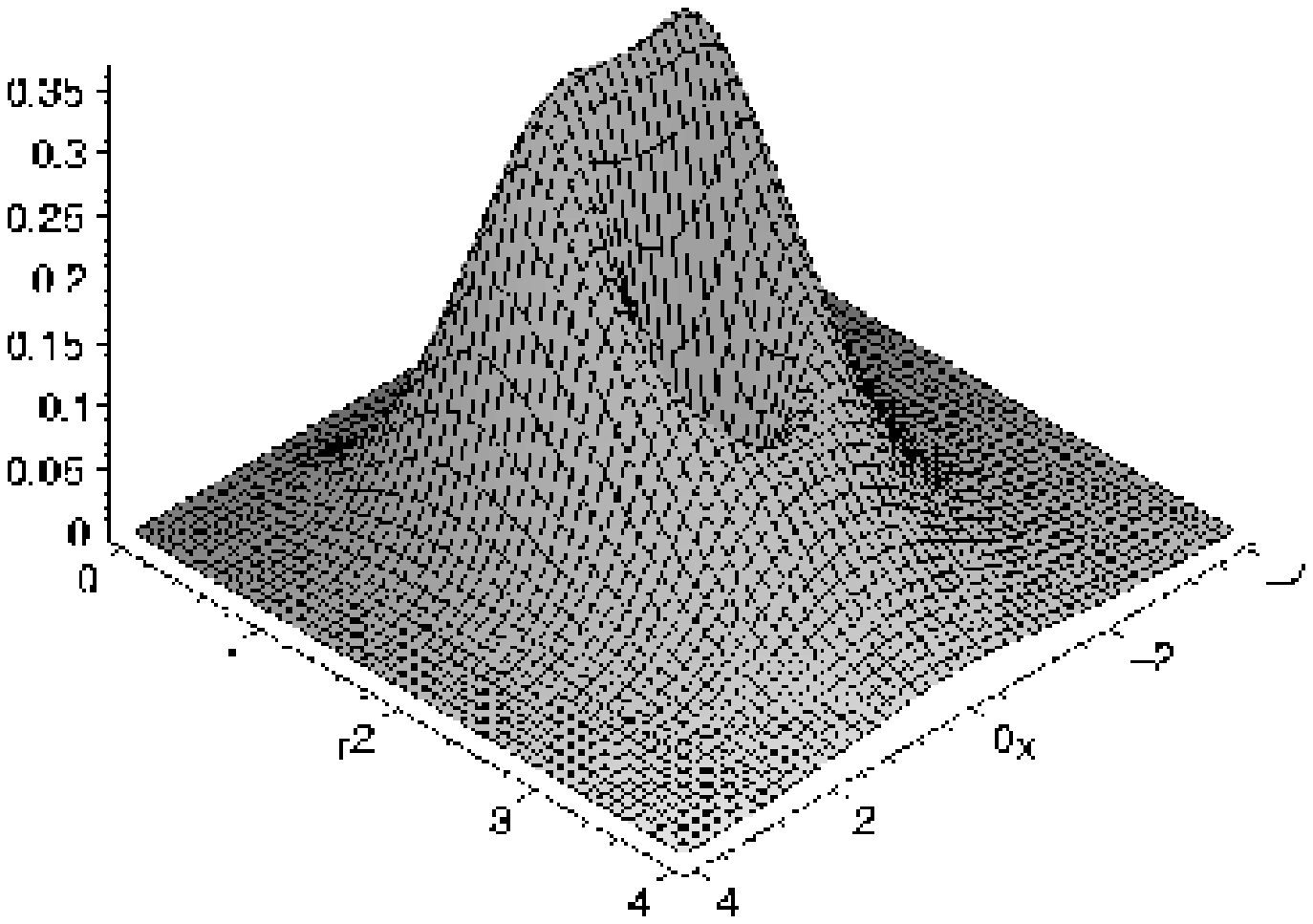}
\put(-260,160){$t=2$}
\vskip -0.5cm
\caption{A stationary soliton configuration at different times.}
\label{fig-station}
\end{figure}

\begin{figure}
\hskip .2cm
\hskip 0.6cm
\epsfxsize=7cm\epsfysize=7cm\epsffile{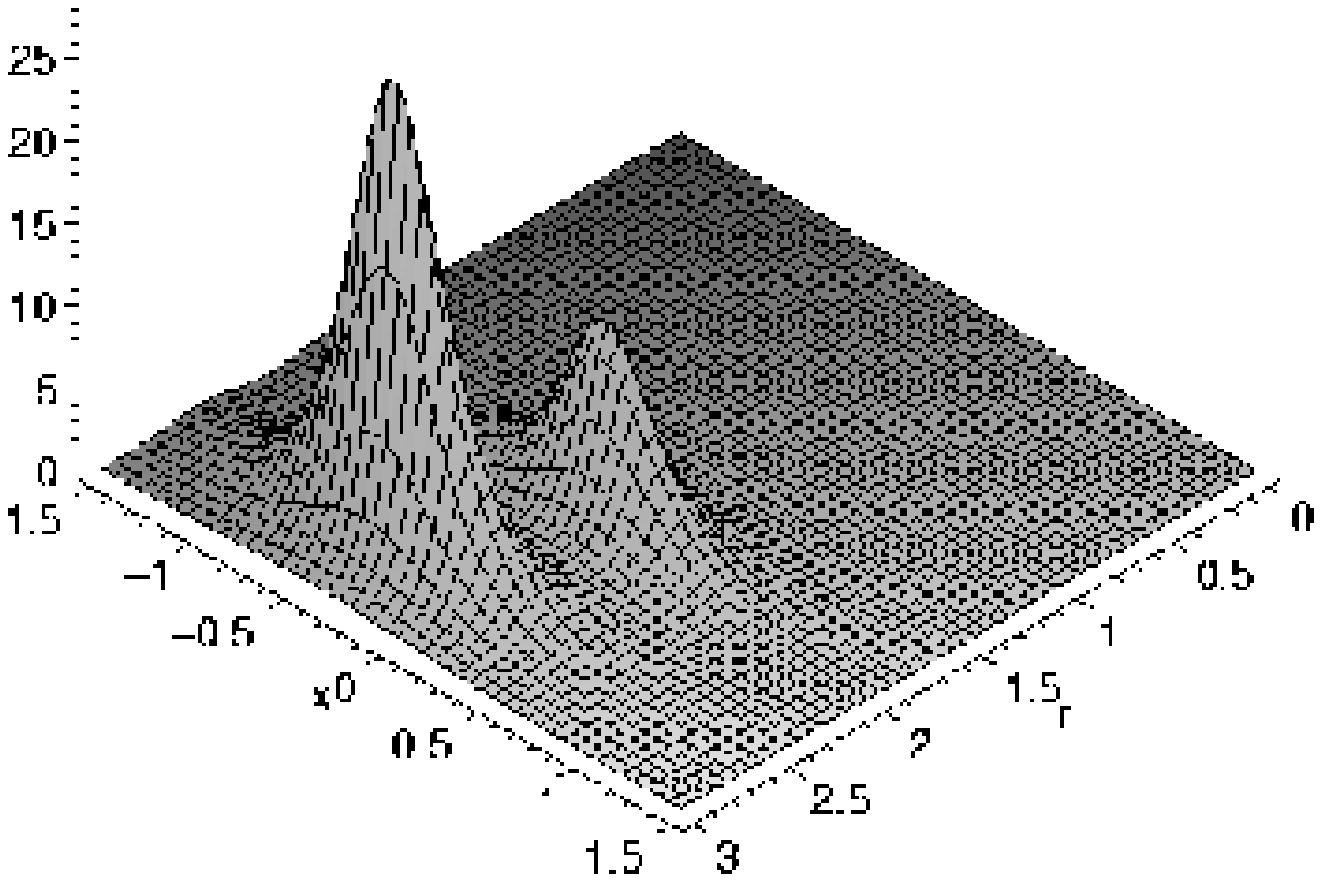}
\put(-250,160){$t=-2.15$}
\par
\vskip -2.5cm
\hskip 6.55cm
\epsfxsize=7cm\epsfysize=7cm\epsffile{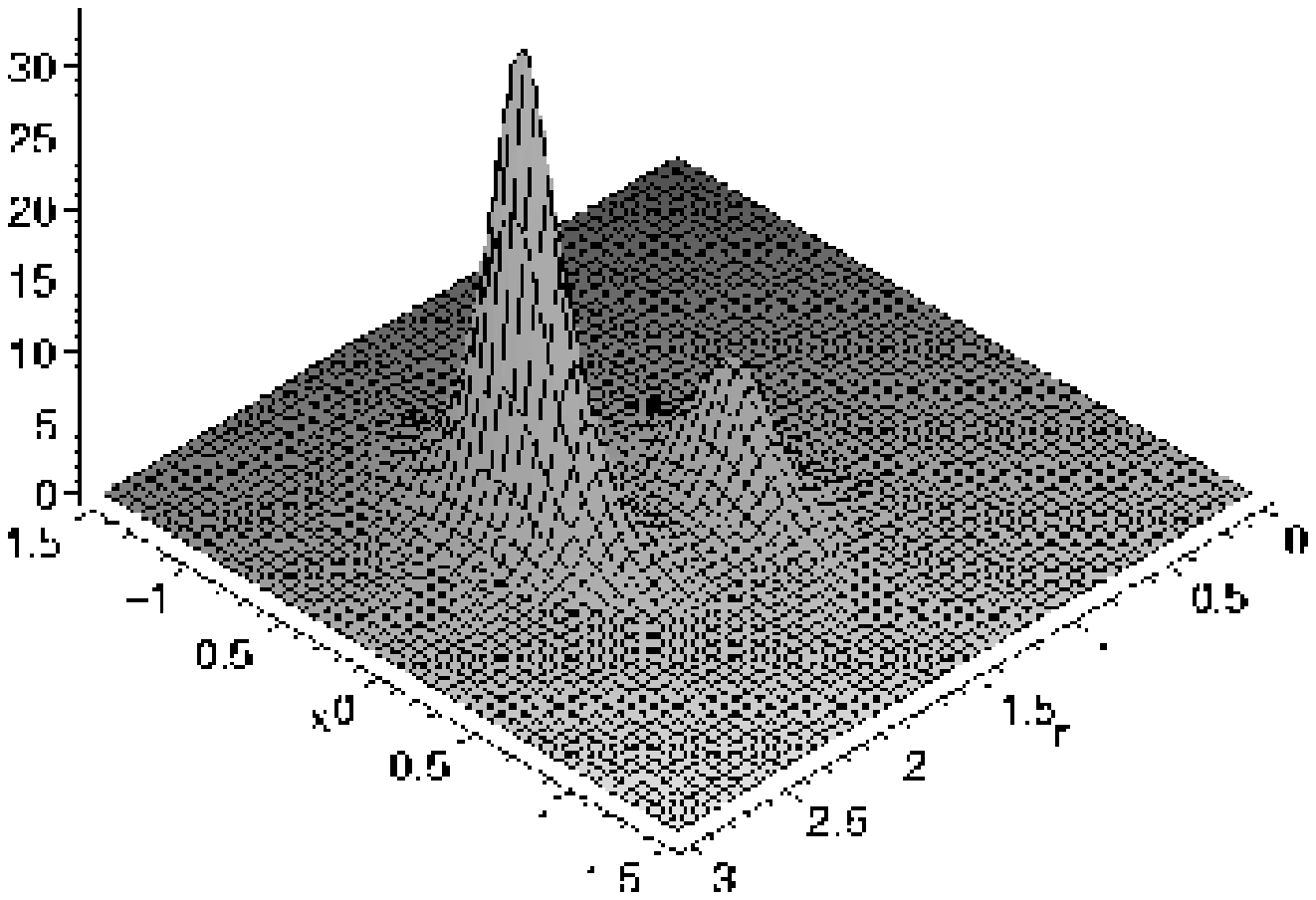}
\put(-240,160){$t=-1.5$}
\par
\vskip -3cm
\hskip 0.2cm
\hskip 0.6cm
\epsfxsize=7cm\epsfysize=7cm\epsffile{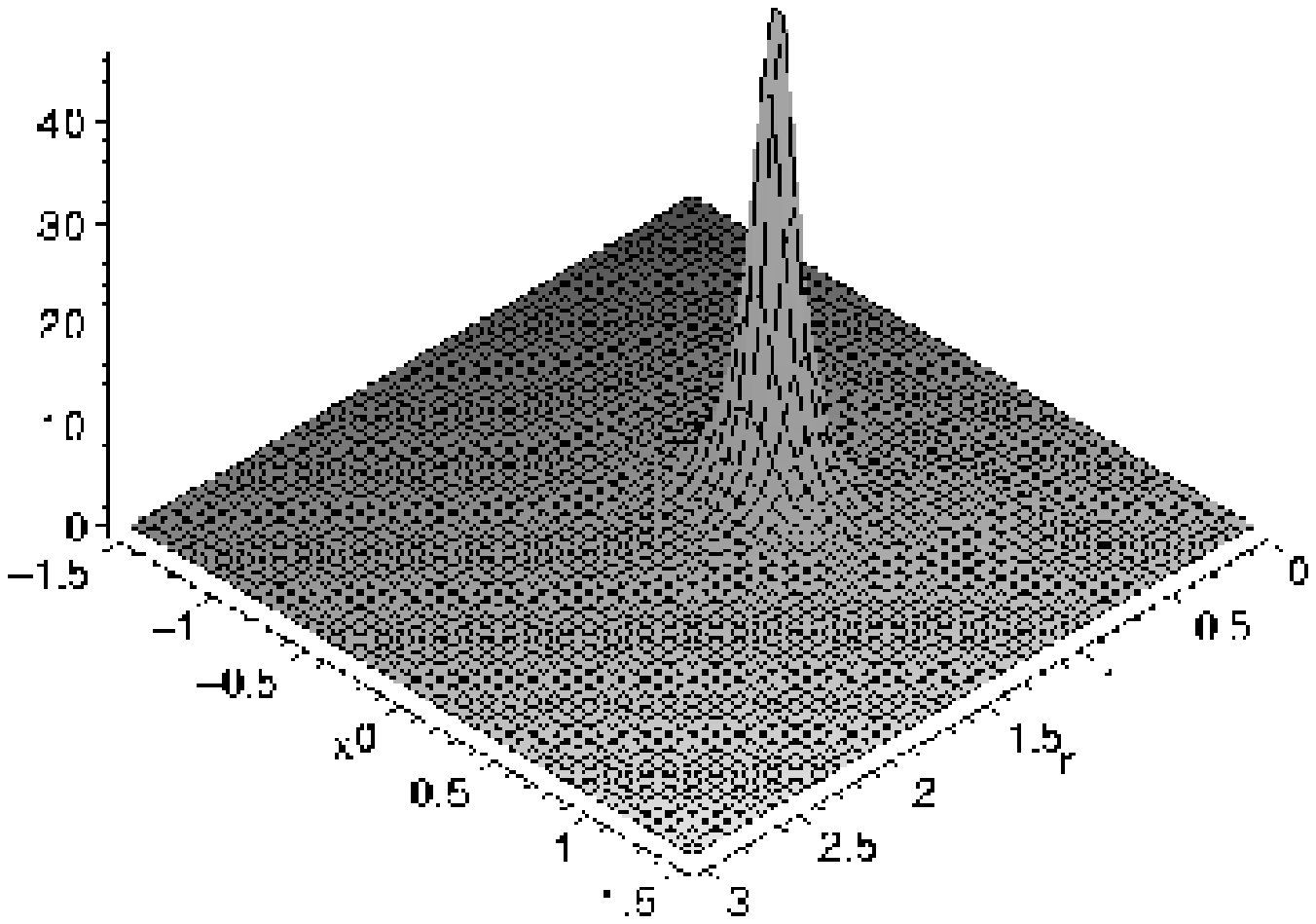}
\put(-250,160){$t=0$}
\par
\vskip -0.5cm
\hskip 6.55cm
\epsfxsize=7cm\epsfysize=7cm\epsffile{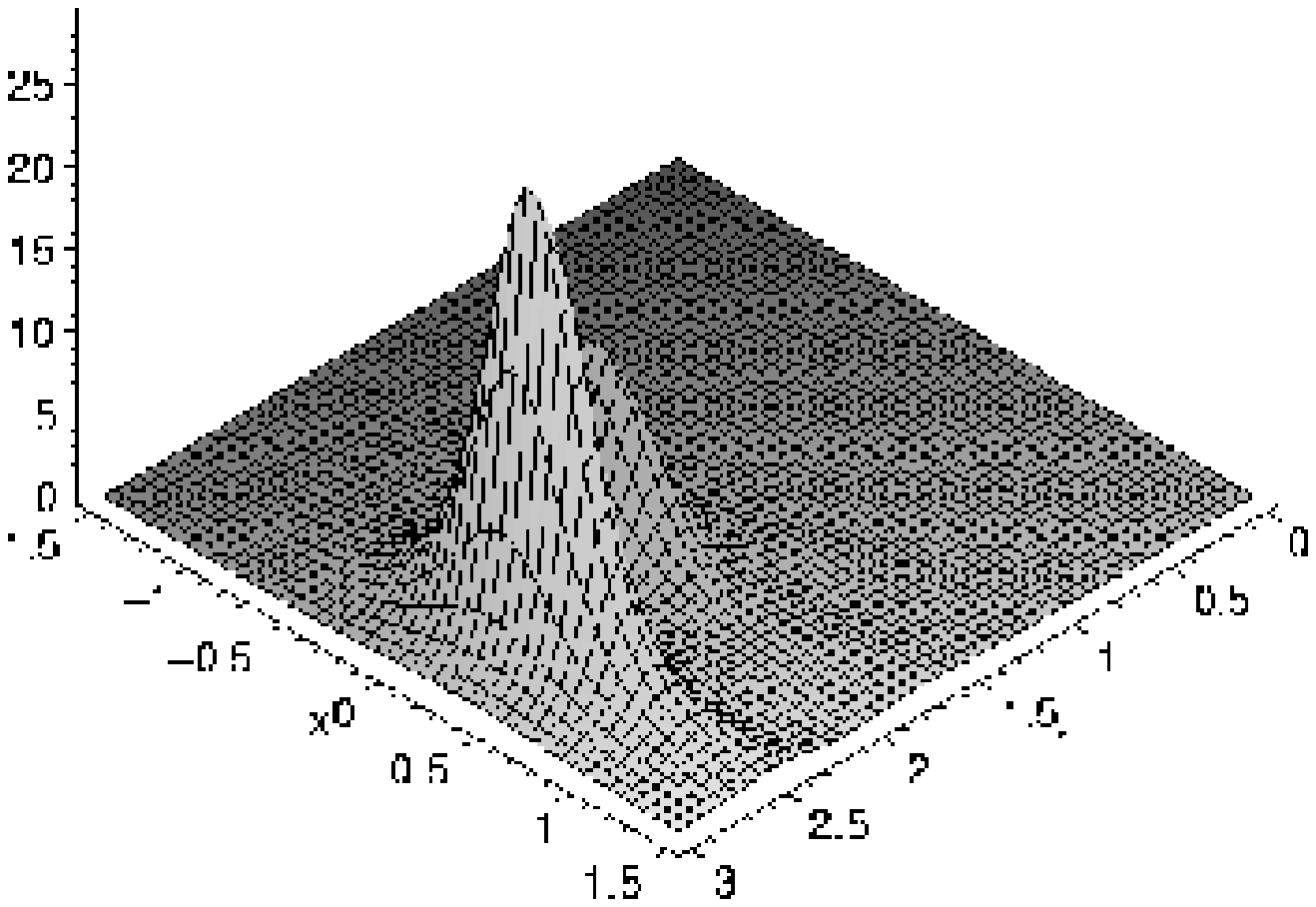}
\put(-240,160){$t=2.15$}
\vskip -.5cm
\caption{A non-trivial dynamics of two solitons in anti-de Sitter space.}
\label{fig-scat}
\end{figure}
\item
Next, we investigate the solution which corresponds to a
non-trivial soliton scattering as presented in Figure
\ref{fig-scat}. The solution has been obtained by
(\ref{Psi-sca}-\ref{pq})
 for the following values of the parameters: $\mu=i$, $f(\omega)=\omega$
and $h(\omega)=\omega^4$ while the picture consists of two
different-sized solitons.
 For large (negative) $t$, the gauge
quantity $-\mbox{tr} \Phi^2$ is peaked at two points, forms a lump
at $t=0$ and then two solitons emerge, for large (positive) $t$,
with the small one been shifted to the left.
\end{itemize}

These solutions can also been obtained using
Uhlenbeck's construction  \cite{Uhl}.
In this approach the function $\Psi$ is assumed to be a product
 of factors  which when substituted in the Lax pair (\ref{Lax})
  the problem simplifies to finding
  solutions of simple first-order partial differential equations \cite{IZ}.
Both methods can be extended to derive solutions  where the function $\Psi$
has higher order pole in $\lm$ (and no others).
Then, $\Psi$ can be written as a product of three (or more) factors with three
(or more) arbitrary vectors (for more details, see \cite{Ioan}).\\

$ \diamondsuit$\, {\it Case where $\bar{\mu}_k=\mu_l$}\\

Secondly, let us construct a large family of solutions which
correspond to the case where $\bar{\mu}_k=\mu_l$.
One way of proceeding is to take the solution $(\ref{Psi})$ with $n=2$,
put $\mu_1=\mu+\varepsilon$, $\mu_2=\bar{\mu}-\varepsilon$ and take the limit
$\varepsilon \ra 0$.
In order for the resulting $\Psi$ to be smooth it is necessary to take
$f_1(\omega_1)=f(\omega_1)$, $f_2(\omega_2)=-1/f(\omega_2)-
\varepsilon h(\omega_2)$
 where $f$ and $h$ are rational functions of one variable.
On taking the limit we obtain a solution $\Psi$ of the form
\be
\Psi=I+\fr{n^1 \otimes m^1}{\lm-\mu}+\fr{n^2 \otimes m^2}{\lm-\bar{\mu}}
\label{sol-ant}
\ee
where $n^k$, $m^k$ for $k=1,2$ are complex valued two vector functions
of the form
\bea
&&\hskip -1.2cm m^1=(1,f),\hs  m^2=(-\bar{f},1)\nonumber\acc
\hskip -0cm\pmatrix{\!n^1\cr n^2\!}\!\!&=&\!\!
\fr{2(\mu-\bar{\mu})}{4(1\!+\!|f|^2)^2\!-\!(\mu\!-\!\bar{\mu})^2|w|^2}
\!\!\pmatrix{\!2(1\!+\!|f|^2) \!&\! -(\mu\!-\!\bar{\mu}) \bar{w}\cr
\!\!(\mu\!-\!\bar{\mu})w \!&\! -2(1\!+\!|f|^2)\!}
\pmatrix{\!m^{1 \dagger} \cr m^{2 \dagger}\!}
\eea
with
\be
w\equiv \fr{2r^2}{(\mu-u)^2}f'+\bar{h}f^2.
\ee
So we generate a solution which depends on the parameter $\mu$ and the
two arbitrary rational functions $f=f(\omega)$ and $h=h(\bar{\omega})$.

\begin{figure}
\hskip .2cm
\epsfxsize=8cm\epsfysize=8cm\epsffile{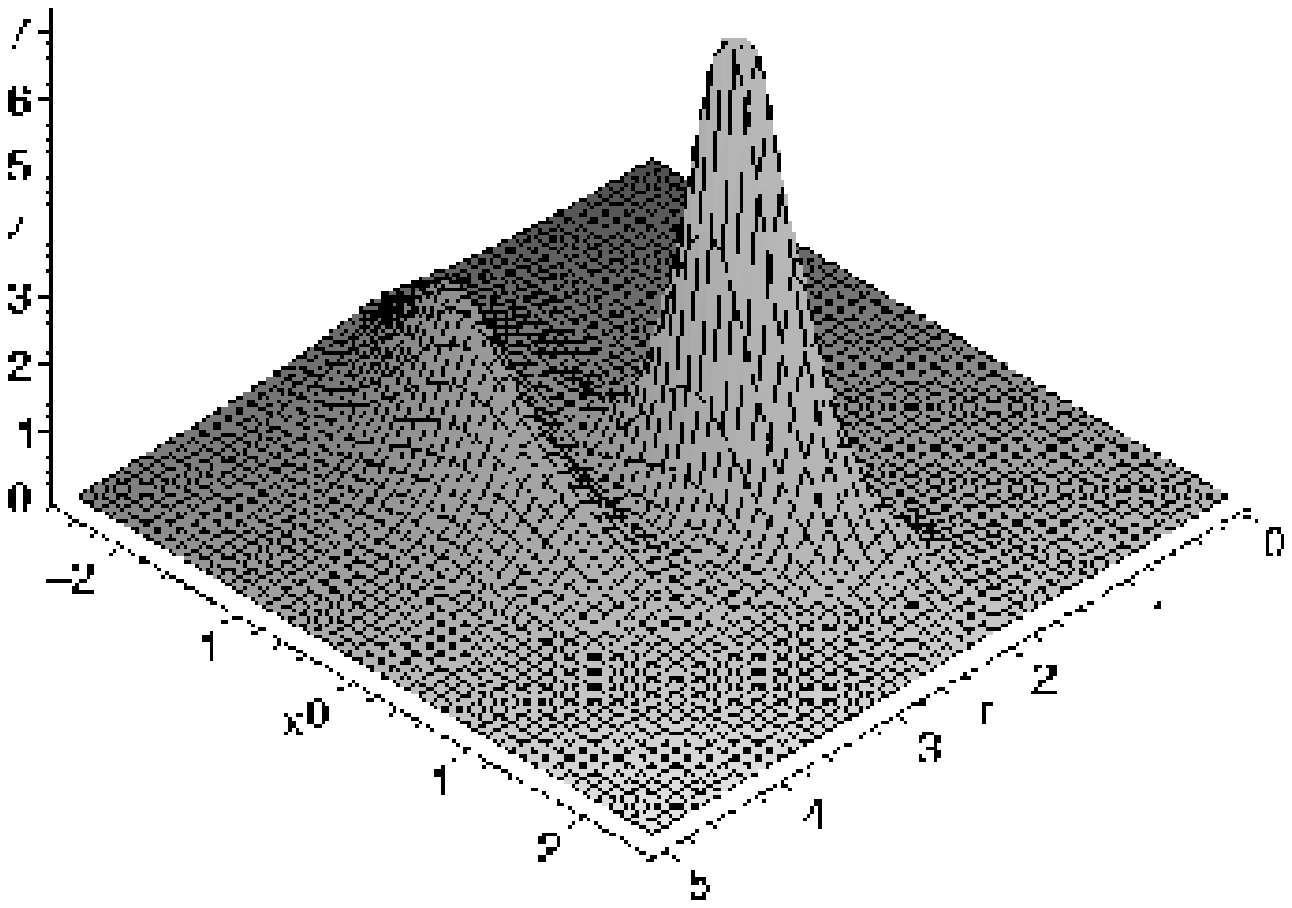}
\par
\vskip -2.8cm
\hskip 7cm
\epsfxsize=8cm\epsfysize=8cm\epsffile{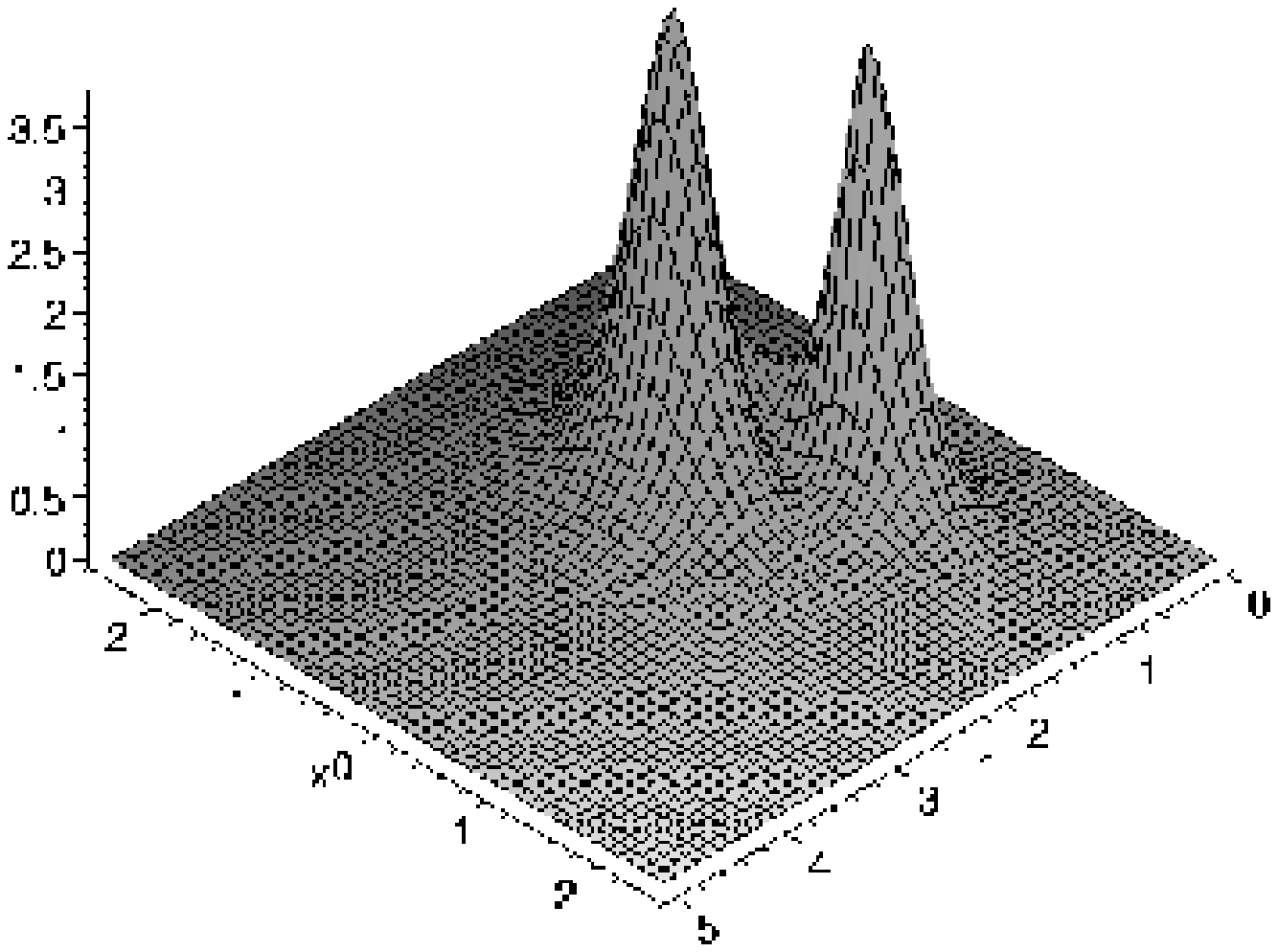}
\par
\hskip .2cm
\vskip -5.5cm
\epsfxsize=8cm\epsfysize=8cm\epsffile{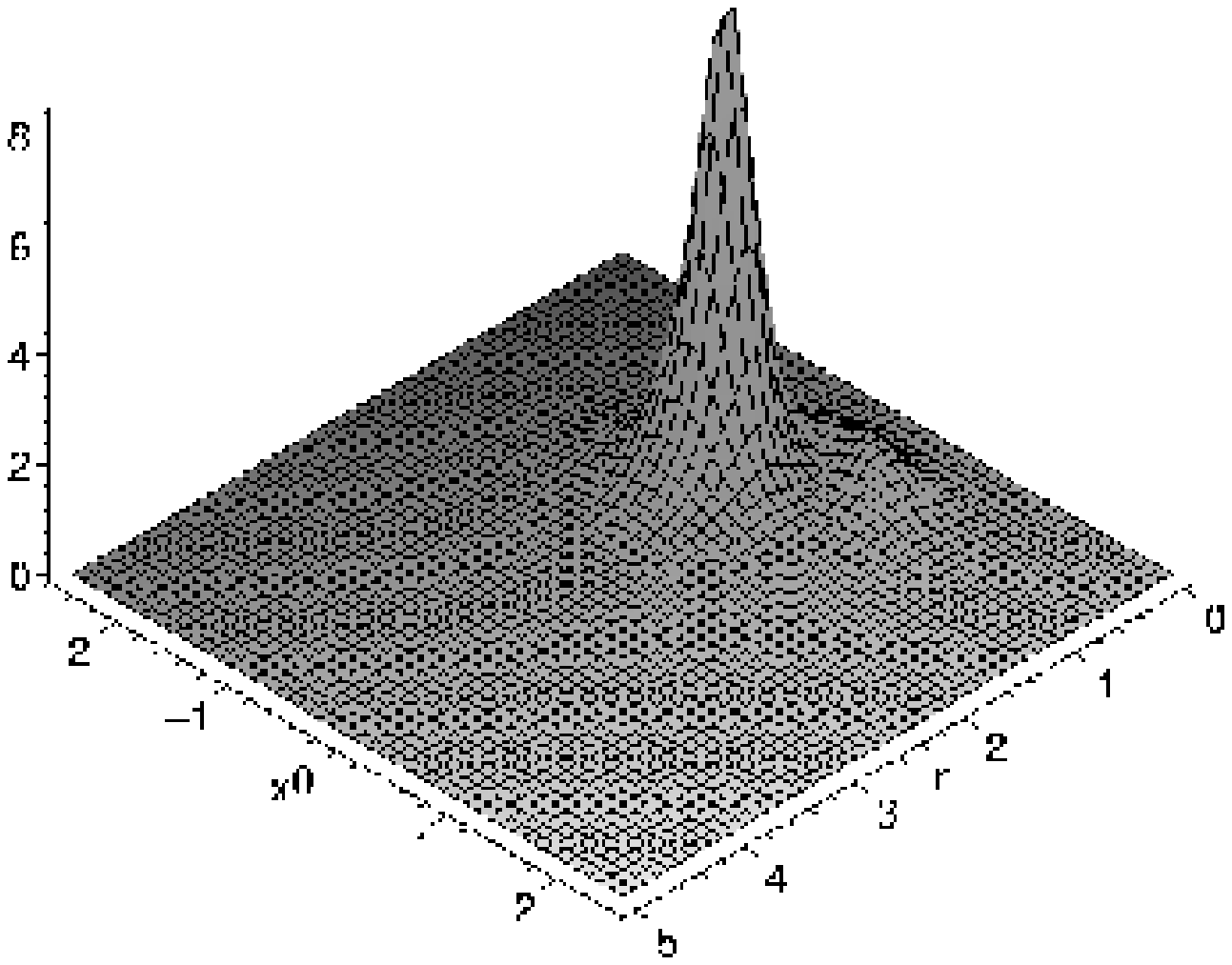}
\par
\vskip -3.8cm
\hskip 6.7cm
\epsfxsize=8cm\epsfysize=8cm\epsffile{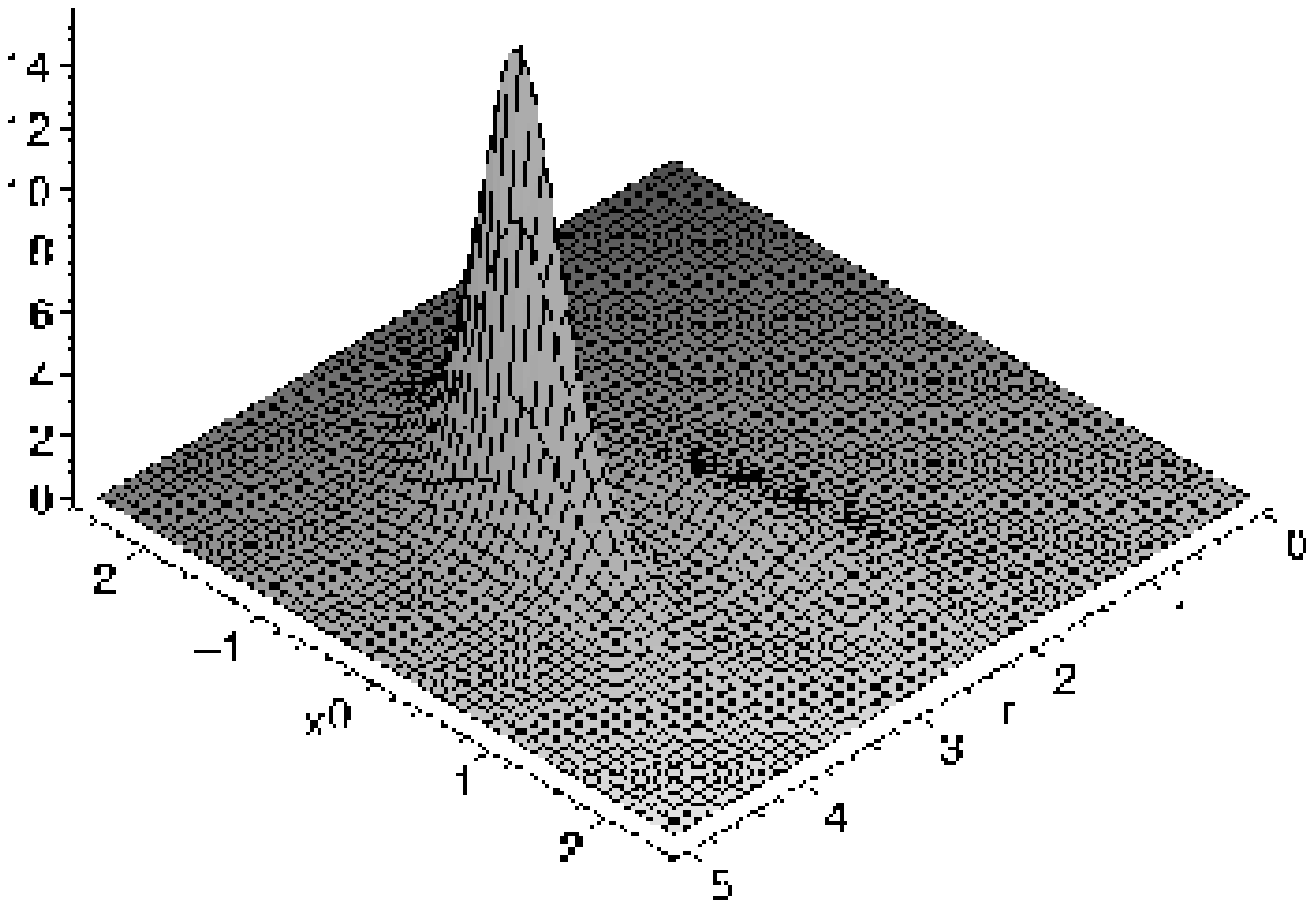}
\vskip -0cm
\caption{A non-trivial dynamics of two solitons in anti-de Sitter space.}
\label{fig-sol-ant}
\end{figure}

In Figure \ref{fig-sol-ant} we represent snapshots of the solution
(\ref{sol-ant}) for $\mu=i$, $f=\omega$ and $h=\bar{\omega}$. The
configuration consists of two solitons with non-trivial scattering
behavior. Again, the quantity $-\mbox{tr} \Phi^2$ is peaked at two
points, for (negative) $t$, which are still distinct at $t=0$ and
then two shifted (compared to the initial ones at $t=-2.5$)
solitons emerge,
 for (positive) $t$.
Throughout the time-evolution their sizes change.
Note that, the scattering solutions belong to a large family
since $f$ and $h$ can be taken to be any rational
functions of $\omega$.

{\it Remark:}
The extension of  the  obtained classical solutions in the whole
anti-de Sitter space, ie using the coordinates $(\rho, \theta, \phi)$,
is unambiguous.
For example, the simplest solution which corresponds to the one soliton
(first derived in \cite{W1}) given by (\ref{Psi}) for $n=1$,
$\mu_1=i$ and $f_1=\omega_1$ implies that
\bea
-\mbox{tr}\,\Phi^2&=&\fr{8r^4}{[(r^2+x^2-t^2)^2+2x^2+2t^2+1]^2}\nonumber\\
&=&\fr{2 \cos^4\rho}{(\cos^2\rho-2)^2}
\eea
which means that  the positive definite quantity $-\mbox{tr}\,\Phi^2$
 is independent of the variables $(\theta,\phi)$.

\section{The Minkowski Model}

The integrable $SU(2)$ chiral model introduced by Ward \cite{W} is
described by the field equation \be \pr_v\left(J^{-1}\pr_u
J\right)-\pr_x\left(J^{-1} \pr_xJ\right)=0 \label{ch} \ee where
the chiral field takes values in: $J \in SU(2)$ while
$u=\fr{1}{2}(t+y)$ and $v=\fr{1}{2}(t-y)$. Note that the
integrable chiral equation is related to the Jarvis equation
studied in \cite{IS}. Once more the equation can be obtained from
the self-dual Yang-Mills-Higgs equation given by (\ref{Bog}) for
specific gauge choice (further details in \cite{Ja}). Then for
specific  boundary conditions spherical symmetric monopole
solutions could be derived using the harmonic map ansatz as
studied in \cite{IS}.

Here the choice of the boundary conditions follows from the
chiral-equation form rather than the gauge-theory form, ie \be
J=J_0+\fr{1}{r}J_1(\theta)+O\left(\fr{1}{r^2}\right), \hs r\ra
\infty \ee for $x+iy=r e^{i\theta}$. Here $J_0$ is a constant
matrix and $J_1$ depends only on $\theta$ (no time dependence).

The model possess a conserved energy density given by \cite{W}
\be
{\cal E}=-\fr{1}{2} \mbox{tr} \left(\left(J^{-1}\pr_t J\right)^2
+\left(J^{-1}\pr_x J\right)^2+\left(J^{-1}\pr_y J\right)^2\right)
\ee
which is a positive-defined functional of the chiral field.

The integrable equation (\ref{ch}) can be written as the compatibility
condition of the system
\bea
\left(\lm \pr_x-\pr_u\right)\psi&=&A\psi\nonumber\\
\left(\lm \pr_v-\pr_x\right)\psi &=&B\psi \label{chcom} \eea where
$\lm \in \C$, $A,B$ are $2 \times 2$ anti-Hermitian trace-free
matrices depending only on $(u,v,x)$ and $\psi(\lm,u,v,x)$ is an
unimodular
 $2 \times 2$ matrix satisfying the reality condition:
$\psi(\lm) \psi(\bar{\lm})^\dg=I$.
The gauge choice of the matrices $A,B$ can be obtained by setting for
$\lm=0$: $\psi(\lm=0)=J(u,v,x)^{-1}$.
Then the system (\ref{chcom}) implies that
\be
A=J^{-1}\pr_uJ,\hs \hs \hs \hs B=J^{-1}\pr_x J.
\ee
Using the standard method of Riemann-Hilbert problem with zeros, Ward
\cite{W} was able to construct the multi-soliton solutions of
(\ref{ch}) by assuming that $\psi$ has a simple poles in $\lm$ and no
others, ie:
\be
\psi=I+\sum_{k=1}^{n} \fr{M_k}{\lm-\mu_k}
\label{cps}
\ee
where $M_k$ are $2\times 2$ matrices independent of the spectral
parameter $\lm$ given by (\ref{M}-\ref{G}), $n$ is the soliton number and
the complex parameter $\mu_k$ defines the velocity of the $k$-th
soliton.
Note that the construction is similar to the construction of baby monopoles
presented in section 3.1.
Again, here the components of $M_k$ are given in terms of a rational
function $f_k$ of the complex variable $\omega_k=x+\mu_k u+\mu_k^{-1}v$.
In particular $m_\al^k$ are holomorphic function of $\omega_k$ defined as:
$m_\al^k=(m_1^k,m_2^k)=(1,f_k)$.
By letting $f_k$ to be exponential functions of $\omega_k$ the
solutions  correspond to extended waves which suffer a phase shift upon
scattering studied in \cite{Leez}.
In general though, by letting $f_k$ to be any rational function of
$\omega_k$ the corresponding soliton dynamics is trivial with no change
of direction or phase shift.

{\it Remark:} In \cite{FI} it was shown that when the spectral
theory was applied for (\ref{ch})  the eigenfunction satisfies a
{\it local Riemann-Hilbert problem} given by: \be \psi^-
=\psi^+\left(I- S(x+\lm u+\lm^{-1} v)\right). \ee and the pure
soliton solutions can be obtained by letting $\lm=\bar{\mu}_l$ in
(\ref{cps}) while the components of the $M_k$ matrix can be
obtained by solving simple algebraic matrix equations.

\subsection{Non-trivial Soliton Scattering}

Here we present the way to obtain solitons with non-trivial
scattering properties, similar to the anti-de Sitter ones. Note
that since the space is flat the non-trivial scattering can be
observed and studied in a more clear way. Once more, we
concentrate in  cases where the formula (\ref{cps}) is not
well-defined: ie when $\mu_k$  are not distinct or when
$\bar{\mu}_k=\mu_l$ for any $(k,l)$.\\

$\diamondsuit$ {\it $\pi/N$ Scattering}\\

Assume that $\psi$ has a double pole in $\lm$ and no others: \be
\psi=I+\sum^2_{k=1} \fr{R_k}{(\lm-\mu)^k} \label{cg} \ee where
$R_k$ are $2\times 2$ matrices independent of $\lm$. This
hypothesis can be generalized by taking $\psi$ to have a pole of
order $n$ in $\lm$ (for more details see \cite{Ioan}). Then $\psi$
satisfies the reality condition if and only if is factorized as
\cite{WI}
 \be \psi(\lm)=\left(I-\fr{(\bar{\mu}-\mu)}{(\lm-\mu)}
\fr{q_1^\dg \otimes q_1} {|q|^2}\right)
\left(I-\fr{(\bar{\mu}-\mu)}{(\lm-\mu)} \fr{q_2^\dg \otimes q_2}
{|q|^2}\right) \label{csol} \ee where $q_i$ for $i=1,2$ are
two-dimensional  vectors of the form: \bea
q_1&=&\left(1+|f|^2\right)\left(1,f\right)+
{\cal \theta} \left(\bar{\mu}-\mu\right)\left(\bar{f},-1\right)\nonumber\\
q_2&=&\left(1,f \right). \eea
 Here $f$ is a rational function of
$\omega=x+\mu u+\mu^{-1}v$ and ${\cal
\theta}=\left(u-\mu^{-2}v\right)f'(\omega)+h(\omega)$. So the
solution corresponds to a large family since one may take $f$ and
$h$ to be any rational meromorphic functions of $\omega$. Then the
chiral field defined as: $J=\psi(\lm=0)^{-1}$ which takes the
product form
 \be J=\left(I+\fr{(\bar{\mu}-\mu)}{\mu} \fr{q_2^\dg
\otimes q_2} {|q|^2}\right) \left(I+\fr{(\bar{\mu}-\mu)}{\mu}
\fr{q_1^\dg \otimes q_1} {|q|^2}\right) \ee
 is smooth on
$\R^{2+1}$ (since the vectors $q_i$ are nowhere zero) and
satisfies the boundary condition irrespectively of the choice of
$f$ and $h$.

The solution (\ref{csol}) has been obtained as a limit of the simple-pole
case (\ref{cg}) with $n=2$.
The idea is to take the limit $\mu_k \ra \mu$ by setting
 $\mu_1=\mu+\varepsilon$, $\mu_2=\mu-\varepsilon$ which implies that
the corresponding functions take the form $f_1(\omega_1)=
f(\omega_1)+\varepsilon h(\omega_1)$ and
$f_2(\omega_2)=f(\omega_2)-\varepsilon h(\omega_2)$.
 Note that in
the limit  $\ve \ra 0$ then $f_2(\omega_2)-f_1(\omega_1)\ra 0$
which results the smoothness of $\psi$. The corresponding solitons
are located when the chiral field $J$
 departs from  its asymptotic value $J_0$ which occurs when
 $\theta \ra 0$.

In principal, it is possible  to visualize the emerging soliton
structures when the functions $f$ and $h$ are rational functions
of degree $p,q\in \N$ of the form: $f(\omega)=\omega^p$ and
$h(\omega)=\omega^q$. In fact, for $q>p$ the configuration
consists of $p-1$ static solitons located at the center-of-mass of
the system forming a ring-like structure if more than one;
accompanied by $N=q-p+1$ solitons which (initially)
accelerate towards the one in the middle, scatter at an angle of
$\pi/N$ and (finally) decelerate as they separate. This
follows from the following analytic argument: the field $J$
departs from its asymptotic value $J_0$ when $\theta=\omega^{p-1}
\left(p\left(u-\mu^{-2}v\right)+\omega^N\right) \ra 0$
which holds when either $\omega^{p-1}=0$ or $\omega^N+p
\left(u-\mu^{-2}v\right)=0$. These are approximately the locations
of the solitons in the $xy$-plane.

Let us illustrate the above family of soliton solutions by examining
two simple cases by giving specific values to the arbitrary parameters
$\mu, f(\omega)$ and $h(\omega)$.

\begin{itemize}
\item Let us investigate the simple case where $\mu=i$, $f(\omega)=\omega$
and $h(\omega)=0$. Then the corresponding energy density takes the
simple form \be {\cal
E}=16\fr{r^4+2r^2+4t^2(2r^2+1)+1}{\left(r^4+2r^2+4t^2+1\right)^2}
\ee where $r$ is the polar coordinate: $r=\sqrt{x^2+y^2}$ while
the energy density is time reversible. Figure \ref{fig-e1}
presents snapshots of the stationary soliton configuration which
forms a single peak at $t=0$ and expands to a ring as $t$
increases.
For large positive $t$, the height of the ring (ie maximum of ${\cal E}$)
is proportional to $t^{-1}$ while its radius is proportional to
$\sqrt{t}$.

\begin{figure}
\hskip .4cm
\epsfxsize=8cm\epsfysize=8cm\epsffile{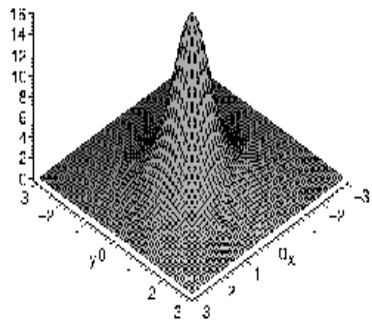}
\put(-280,160){$t=0$}
\par
\vskip -1.25cm
\hskip 5cm
\epsfxsize=8cm\epsfysize=8cm\epsffile{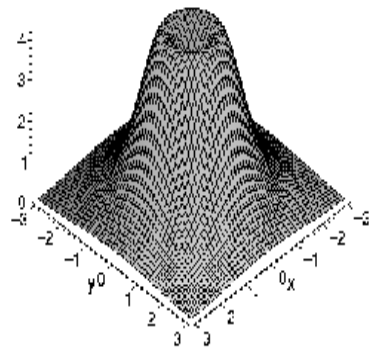}
\put(-260,160){$t=1$}
\vskip -.5cm
\caption{A stationary soliton configuration at different times in flat
spacetime.}
\label{fig-e1}
\end{figure}

\begin{figure}
\hskip .2cm
\epsfxsize=7cm\epsfysize=7cm\epsffile{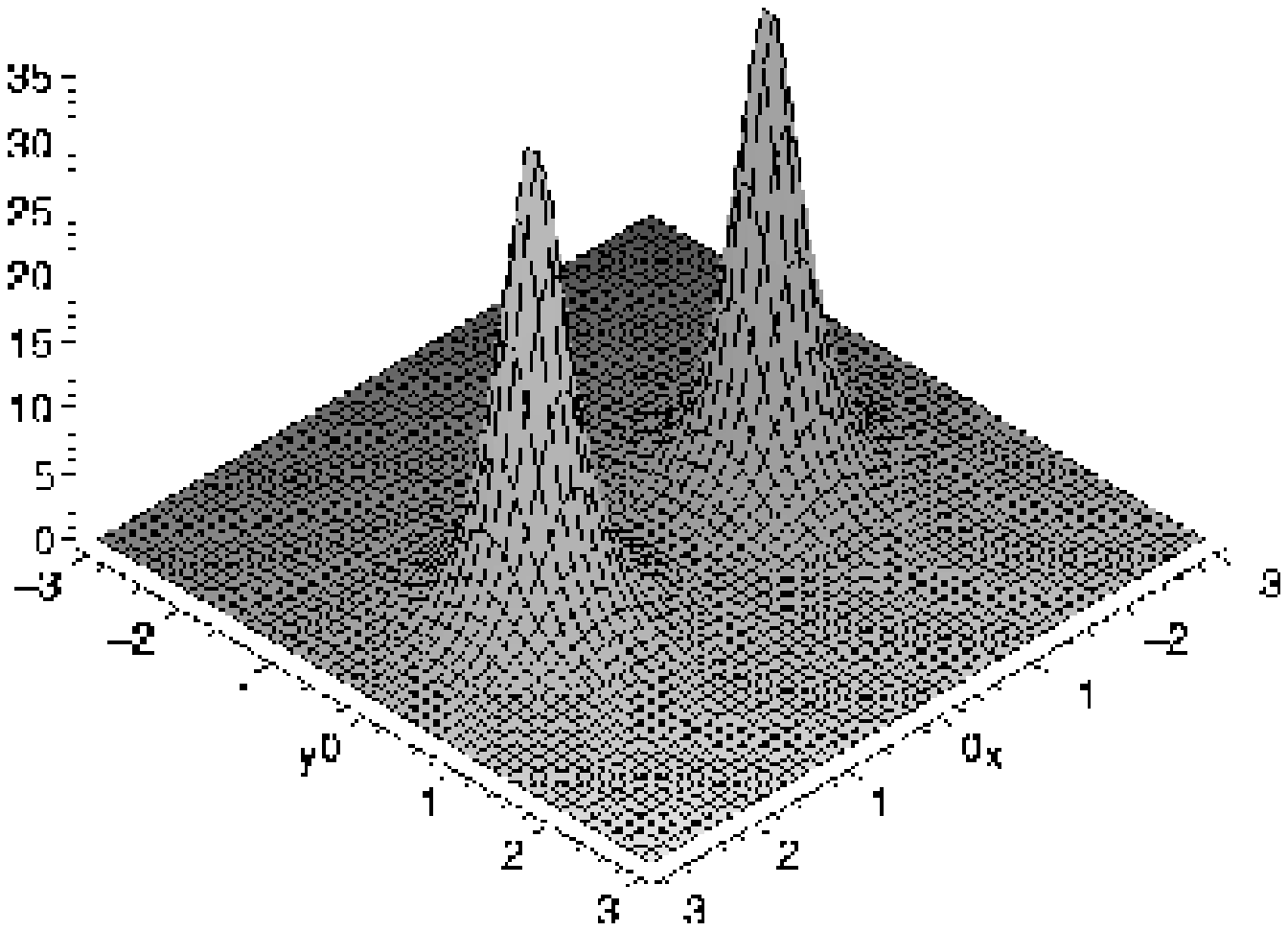}
\par
\vskip -2.25cm
\hskip 6cm
\epsfxsize=7cm\epsfysize=7cm\epsffile{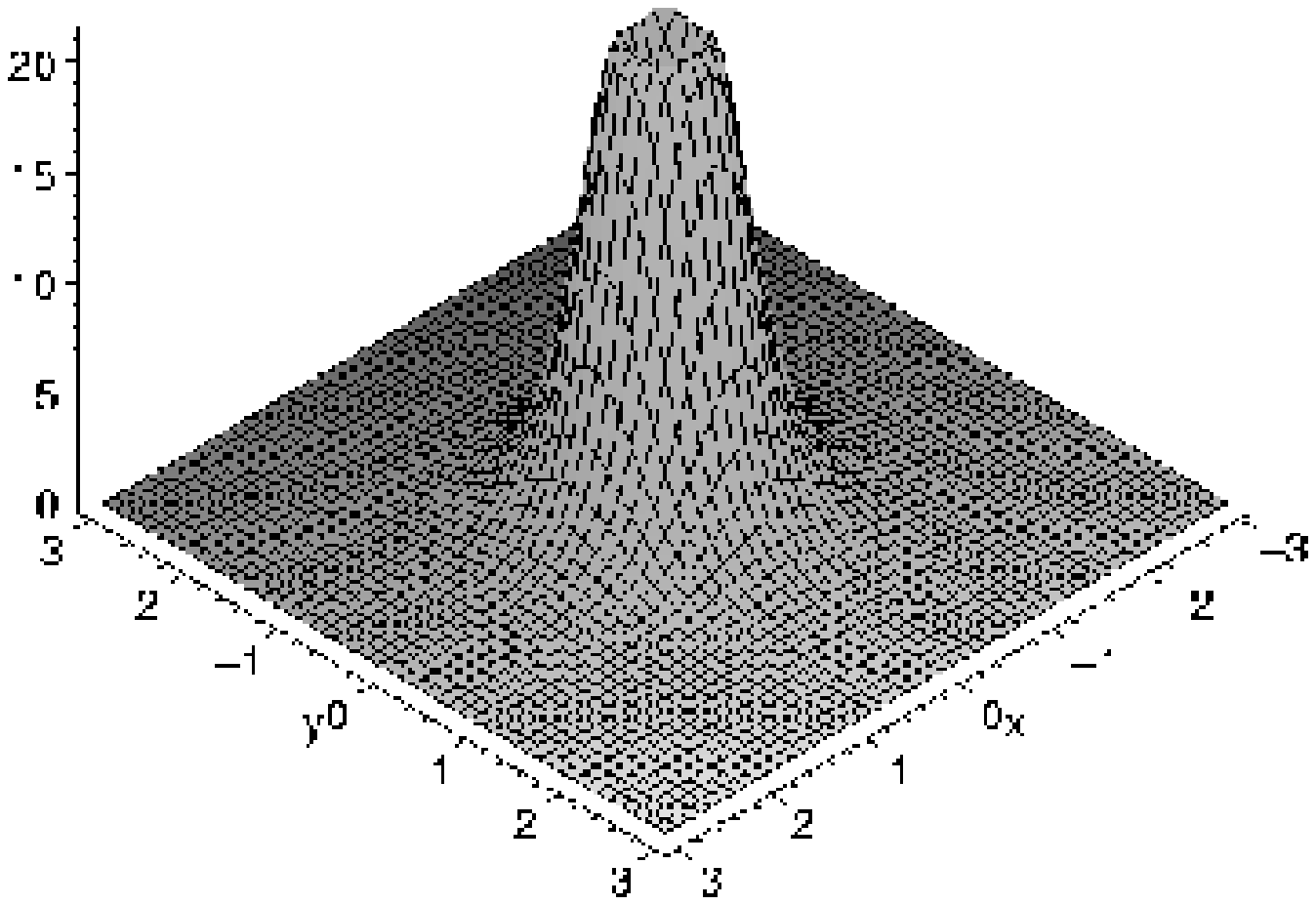}
\par
\hskip .2cm
\vskip -2.5cm
\epsfxsize=7cm\epsfysize=7cm\epsffile{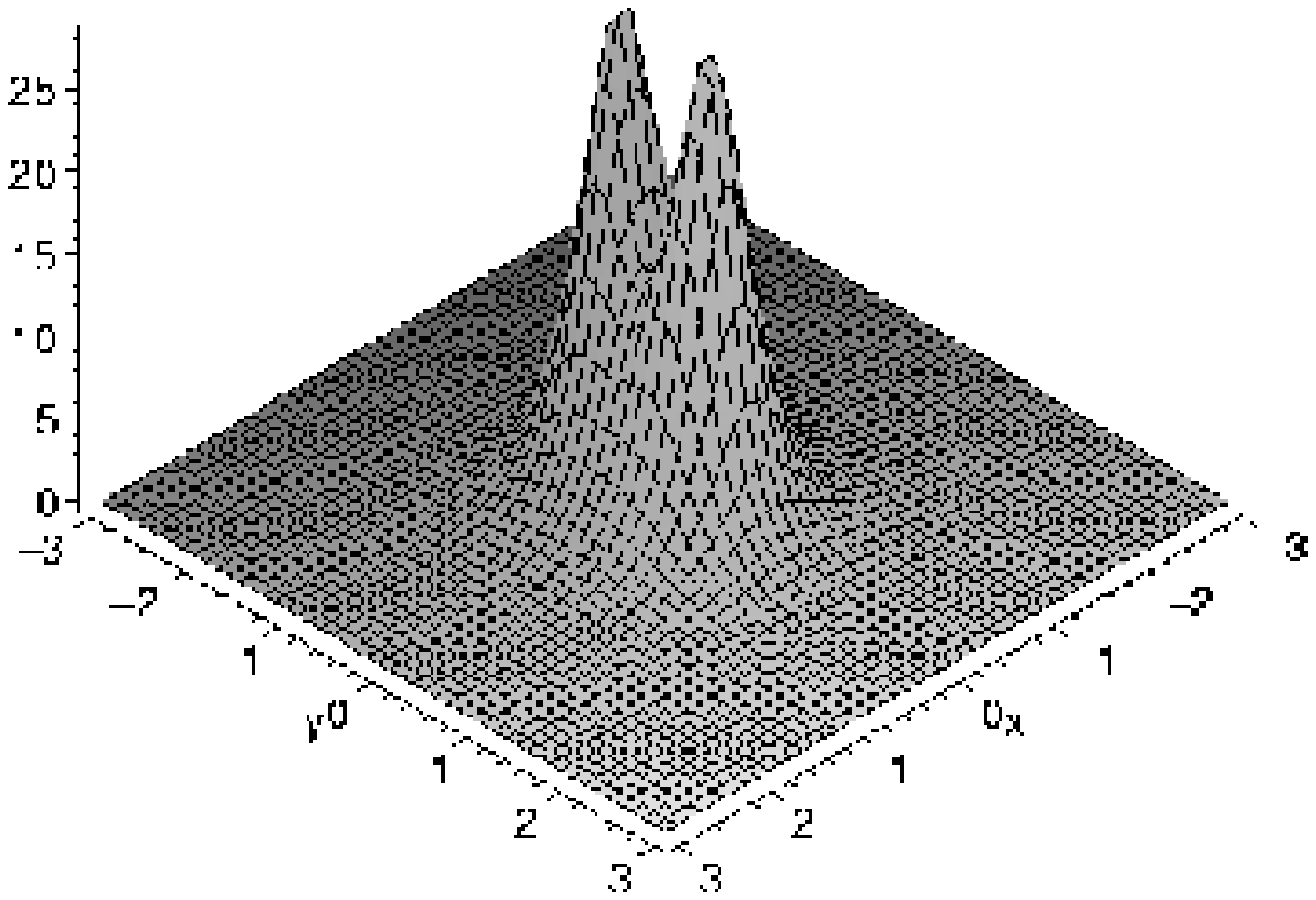}
\par
\vskip -2.25cm
\hskip 6cm
\epsfxsize=8cm\epsfysize=7cm\epsffile{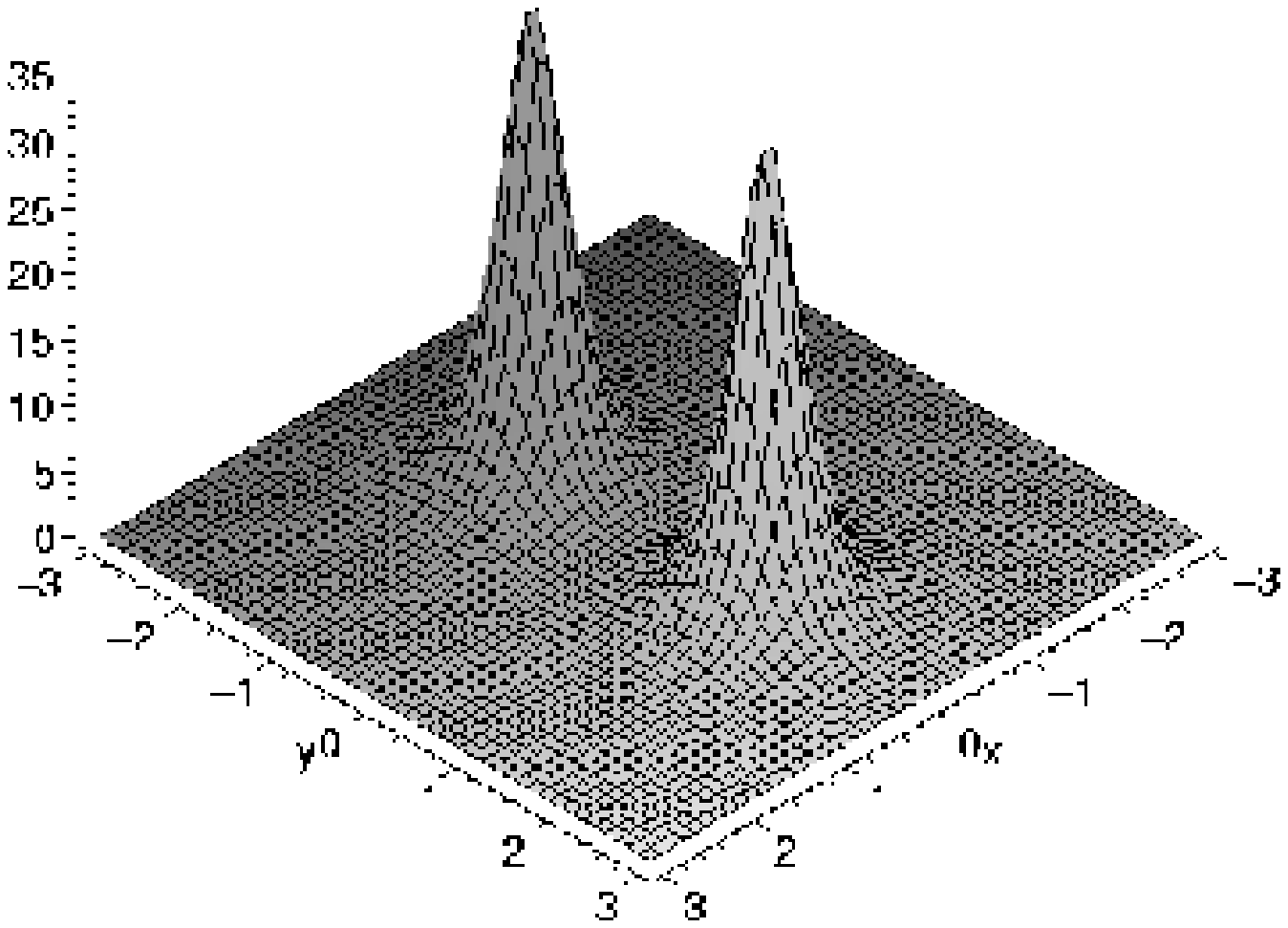}
\vskip -.5cm 
\caption{A $90^0$ scattering
of two soliton in flat spacetime.} \label{fig-es}
\end{figure}

\item Next let us take $\mu=i$, $f(\omega)=\omega$ and $h(\omega)=\omega$.
The energy density is given by \be {\cal E}=16
\fr{5r^4+10r^2+4t^2(1+2r^2)-8t(x^2-y^2)+1}
{\left(5r^4+2r^2+4t^2+8t(x^2-y^2)+1\right)^2} \ee and is symmetric
under the interchange $t \ra -t, x \ra y, y\ra x$. Figure
\ref{fig-es} presents the collision of two solitons near $t=0$:
the solitons accelerate towards each other, scatter at right
angles and decelerate as they separate. In particular, the
collision is time symmetric and elastic (no emitted radiation).

Note that for negative $t$, the two solitons (following the asymptotic
argument of $J$) are located on the $x$-axis at $x \approx \pm \sqrt{t}$
while
for positive $t$, they are located in the $y$-axis at $y \approx
\pm \sqrt{-t}$.
Once more the corresponding solitons are not of constant size:
their height is proportional to $t^{-1}$ while their radii are
proportional to $\sqrt{t}$.
In fact the solitons spread out as they move apart.

\end{itemize}

Although it seems strange that by taking the limit of soliton
solutions with trivial scattering soliton solutions with
non-trivial scattering have been obtained there is an explanation
as shown in \cite{Ioan}. By studying the effect of the variation
of $\varepsilon$ to a two soliton configuration it has been
observed that at the limit $\varepsilon \ra 0$ the solitons
disperse, shift and interact with each other since their internal
degrees of freedom and their impact parameter change. As a result
the two initial well-separated solitons form a ring at the limit
$\varepsilon \ra 0$.

In general in a head-on collision of $N$ indistinguishable
solitons the scattering angle of the emerging structures relative
to the incoming ones can be $\pi/N$. In particular when the $N$
solitons are very close together they merge and form a ring-like
structure and finally they emerge from the ring
in a direction that bisects the angle formed by the incoming ones.\\

$\diamondsuit$ {\it Elastic soliton-antisoliton non-trivial scattering}\\

Next we construct a large family of solutions
which represent soliton-antisoliton field configurations \cite{Ioan}.
The corresponding solution has the form
\be
\psi(\lm)=I+\fr{n^1\otimes m^1}{\lm-i}+\fr{n^2\otimes m^2}{\lm+i}
\ee
where $n^k,m^k$ (for $k=1,2$) are complex-valued two-dimensional vectors
independent of $\lm$ of the form:
\bea
m^1&=&(1,f),\hs \hs m^2=(-\bar{f},1)\nonumber\acc
\pmatrix{n^1\cr n^2}\!\!&=&\!\!
\fr{1}{\left(1+|f|^2\right)^2+|w|^2}
\pmatrix{2i\left(1+|f|^2\right) & 2 \bar{w}\cr
-2w & -2i\left(1+|f|^2\right)}
\pmatrix{m^{1 \dagger} \cr m^{2 \dagger}}
\eea
with
\be
w=\bar{h}f^2+2tf'.
\ee
So we generate a solution $J=\psi(\lm=0)$ which depends on the two
arbitrary rational functions $f=f(z)$ and $h=h(\bar{z})$ where $z=x+iy$.
In the general case where $f=z^p, h=\bar{z}^q$ for $p,q>0$ the
energy of the configuration is equal to: $E=(2p+q)\,8\pi$
 and the corresponding solutions consists of $2p+q$ lumps at
the $xy$-plane which scatter non-trivially and are combinations of
solitons and antisolitons. To prove that the corresponding
configuration consists of solitons and antisolitons a topological
charge was introduced for the integrable chiral model by
exploiting its connections with the $O(3)$ sigma model.

A topological charge may be defined for the chiral model (although
it is not a topological model) by exploiting its connection with
the O(4) sigma model \cite{Leez,si}; ie by letting \be J=I \phi_0+
i\, \sbf \,\mbox{\boldmath $\phi$} \ee where $\sbf$ are the usual Pauli matrices and
$(\phi_0,\mbox{\boldmath $\phi$})= (\phi_0,\phi_1,\phi_2,\phi_3)$ is a four vector of
real fields that are constrained to  lie on $S^3$ with the
constraint $\phi_0^2+\mbox{\boldmath $\phi$} \,\mbox{\boldmath $\phi$}=1$. The only static finite energy
solutions of the O(4) sigma model correspond to the embedding of
the O(3) sigma model \cite{BG}. Therefore the only static finite
energy solutions of (\ref{ch}) are the O(3) embeddings that we
shall describe. This is because for the one-soliton solution
(static or Lorentz boosted in the $y$-axis) the system behaves
like the O(4) model for which the O(3) embedding is totally
geodesic. However for time-dependent configurations the evolution
of the field will not general lie in an O(3) subspace of O(4).

In studying soliton-like solutions we require that the field
configuration has finite energy.
This implies that the field must  take the same value at all points
of spatial infinity so that the space is compactified from $\R^2$ to
$S^2$.
At fixed time the field is a map from $S^2$ into the target space.
Now for the O(3) model the field is a map $\phi: S^2 \rightarrow S^2$
 and due to the homotopy relation
\be \pi_2(S^2)=Z \ee such maps are classified by an integer
winding number ${\cal N}$ which is a conserved topological charge.
An expression for this charge is given by \be {\cal
N}=\frac{1}{8\pi} \int \epsilon_{ij} \, \mbox{\boldmath $\phi$}\left(\partial_i \mbox{\boldmath $\phi$}
\wedge \partial_j  \mbox{\boldmath $\phi$}\right) d^2x \ee where $i=1,2$ while
$x^i=(x,y)$.

On the contrary, for the O(4) sigma model the field at fixed time
is a map $(\phi_0,\mbox{\boldmath $\phi$}): S^2 \rightarrow S^3$ which implies that
 \be \pi_2(S^3)=0 \ee and therefore there is no winding
number. However, for soliton solutions that correspond to some
initial embedding of O(3) space into O(4) there is a useful
quantity presented below.

Consider the O(4) configuration which at some time corresponds to
an O(3) embedding which we choose to be $\phi_0=0$ for
definiteness. At this time the field is restricted to an $S^2$
equator of the possible $S^3$ target space. Suppose that the field
never maps to the antipodal points $(1,-1)$ so the target space is
$S^3_0 \simeq S^2 \times \R$ and thus we have the homotopy
relation
 \be
\pi_2(S_0^3)=\pi_2(S^2\times \R)=\pi_2(S^2) \oplus \pi_2(\R)=Z
 \ee
and therefore a topological winding number exists. An expression
for this winding number is easy to give since it is the winding
number of the map after projection onto the chosen
 $S^2$ equator ie
\be {\cal N'}=\frac{1}{8\pi}\int \epsilon_{ij} \,\mbox{\boldmath $\phi$}'
\left(\partial_i \mbox{\boldmath $\phi$}' \wedge \partial_j \mbox{\boldmath $\phi$}'\right) d^2x \ee
where $\mbox{\boldmath $\phi$}'=\mbox{\boldmath $\phi$}/|\mbox{\boldmath $\phi$}|$. If the field does map to the antipodal
points at some time the winding number is ill defined at this time
and if considered as a function of time ${\cal N'}$ will be
integer valued but may suffer discontinuous jumps as the field
moves through the antipodal points. In the following examples
before comparing the solution $J$ with the O(3) embedding it is
convenient to perform the transformation $J \rightarrow MJ$ with
\be M=\frac{1}{\sqrt{2}} \pmatrix{\,\,\,1 & 1\cr -1 & 1} \ee so
that the evolution of the field remains close to the O(3)
embedding.

One particular example is given below:

\begin{figure}
\hskip .2cm
\epsfxsize=7cm\epsfysize=7cm\epsffile{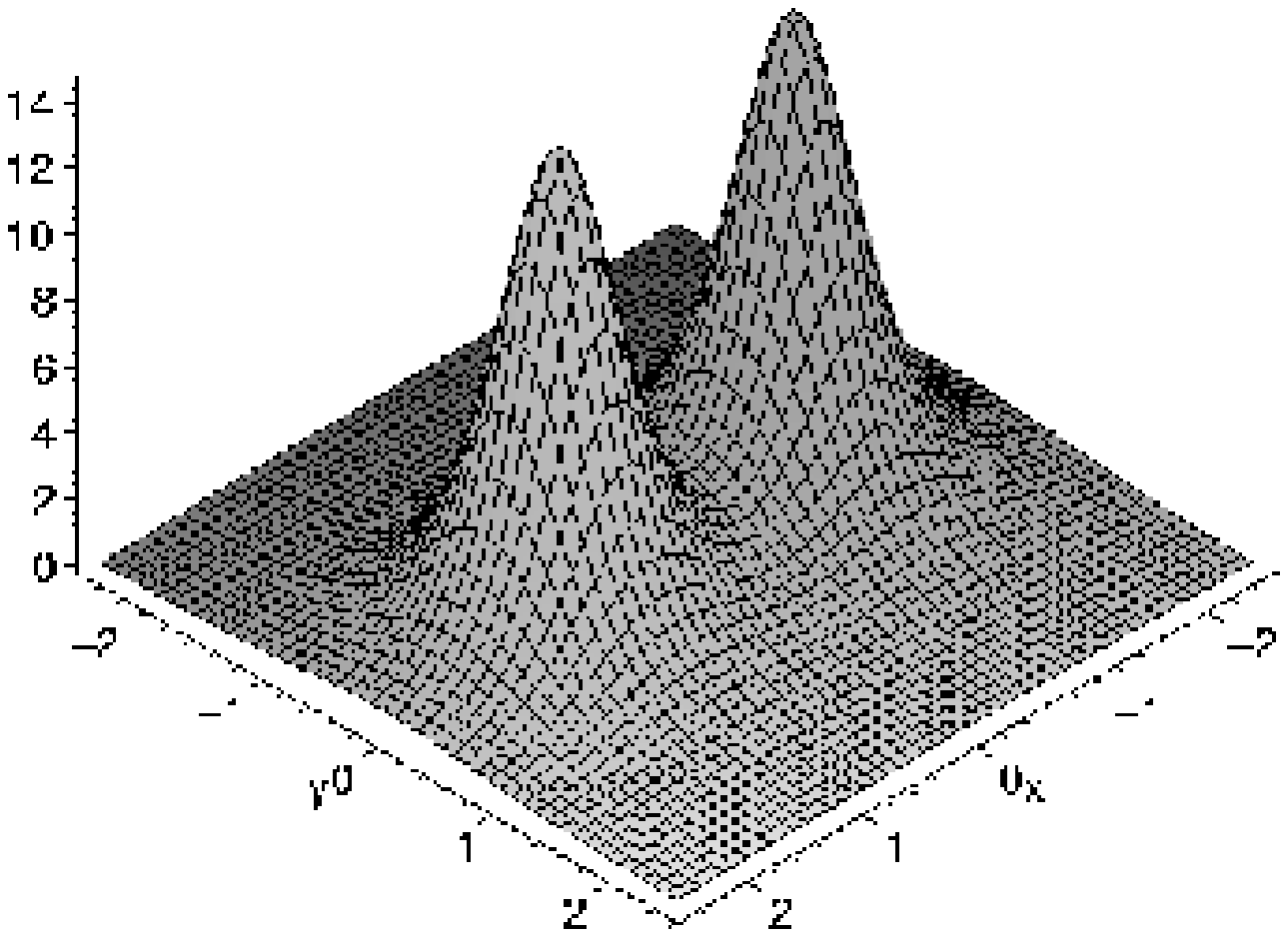}
\put(-250,160){$t=-1$}
\par
\vskip -1.9cm
\hskip 7cm
\epsfxsize=7cm\epsfysize=7cm\epsffile{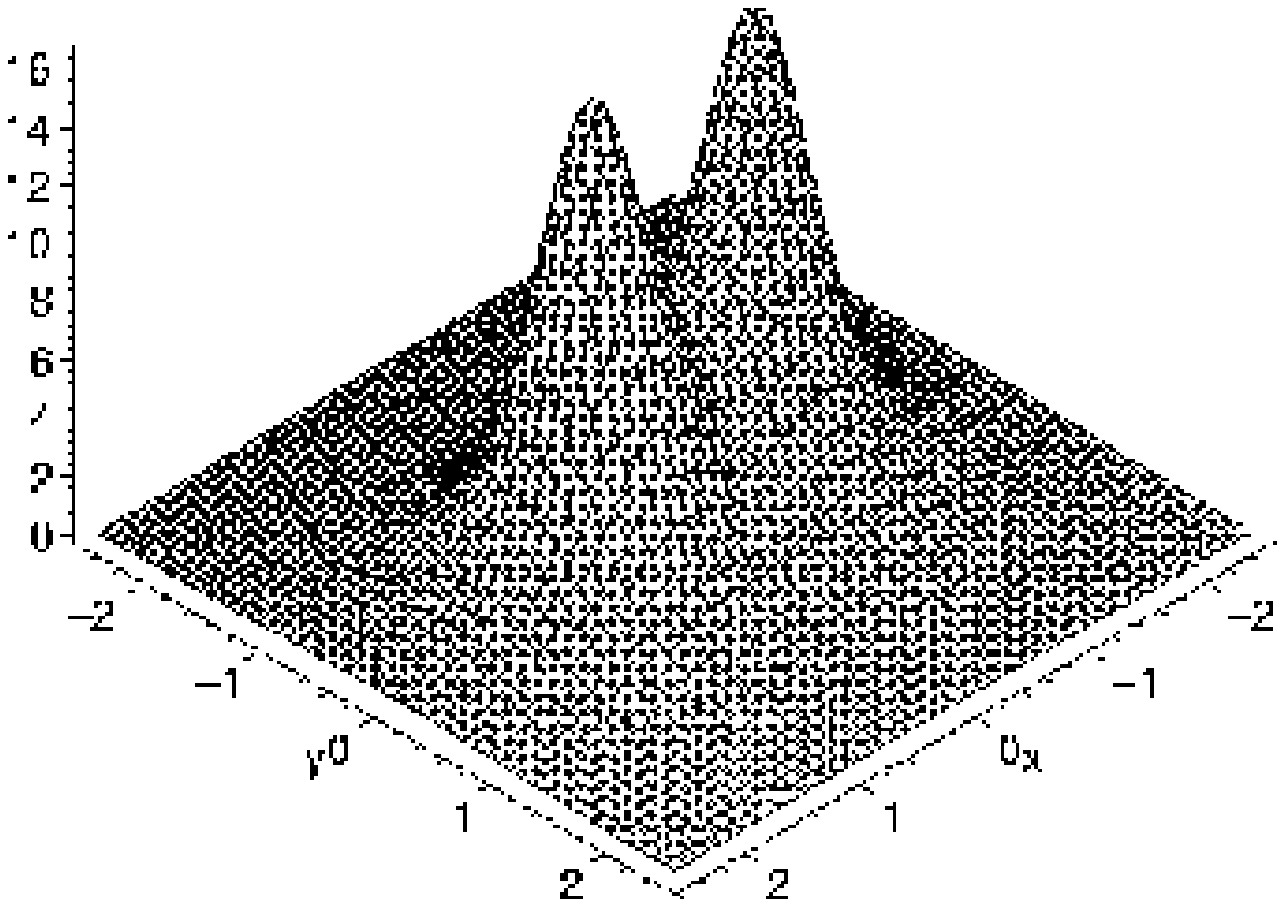}
\put(-240,100){$t=-0.5$}
\par
\vskip -3cm
\hskip .2cm
\epsfxsize=7cm\epsfysize=7cm\epsffile{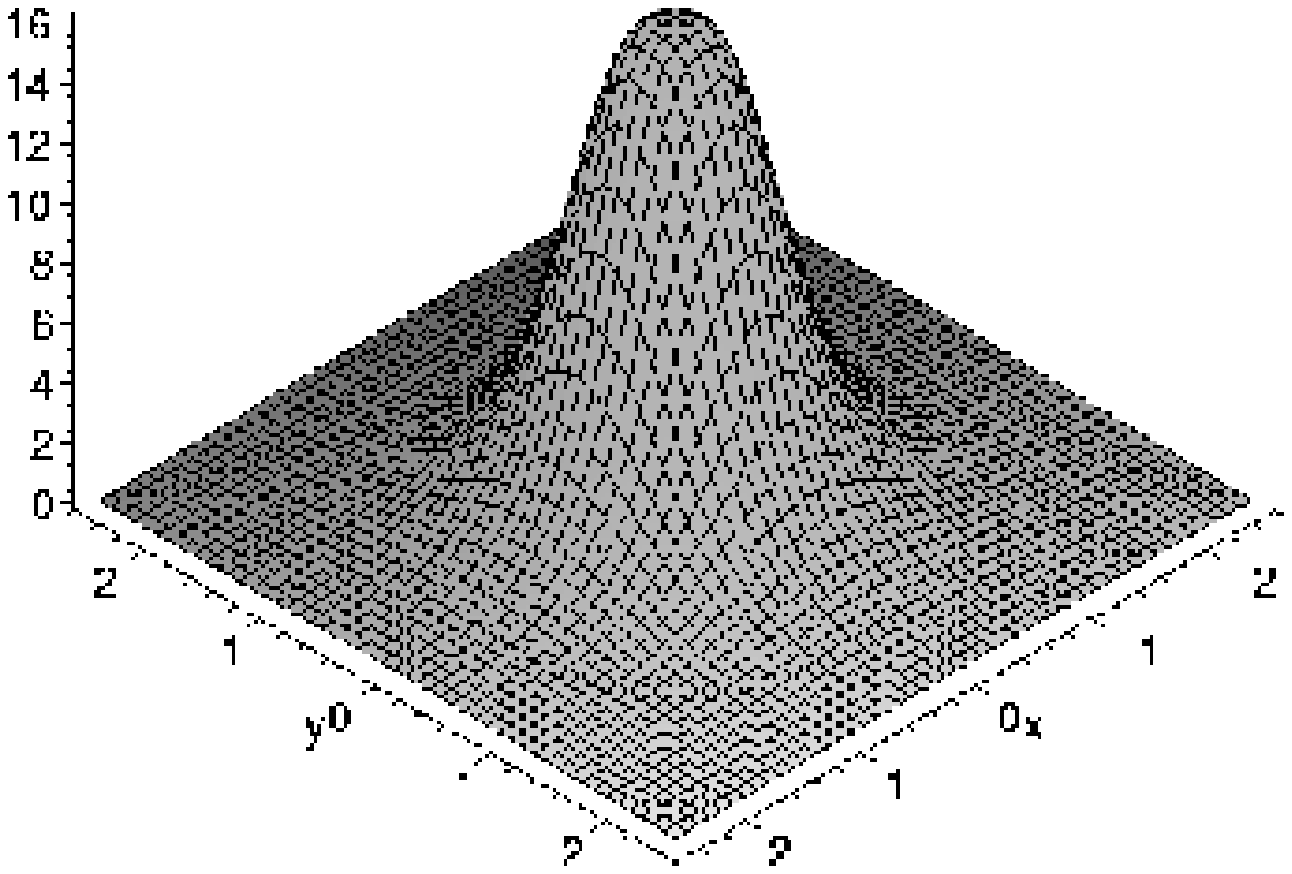}
\put(-250,160){$t=0$}
\par
\vskip -1.8cm
\hskip 7cm
\epsfxsize=7cm\epsfysize=7cm\epsffile{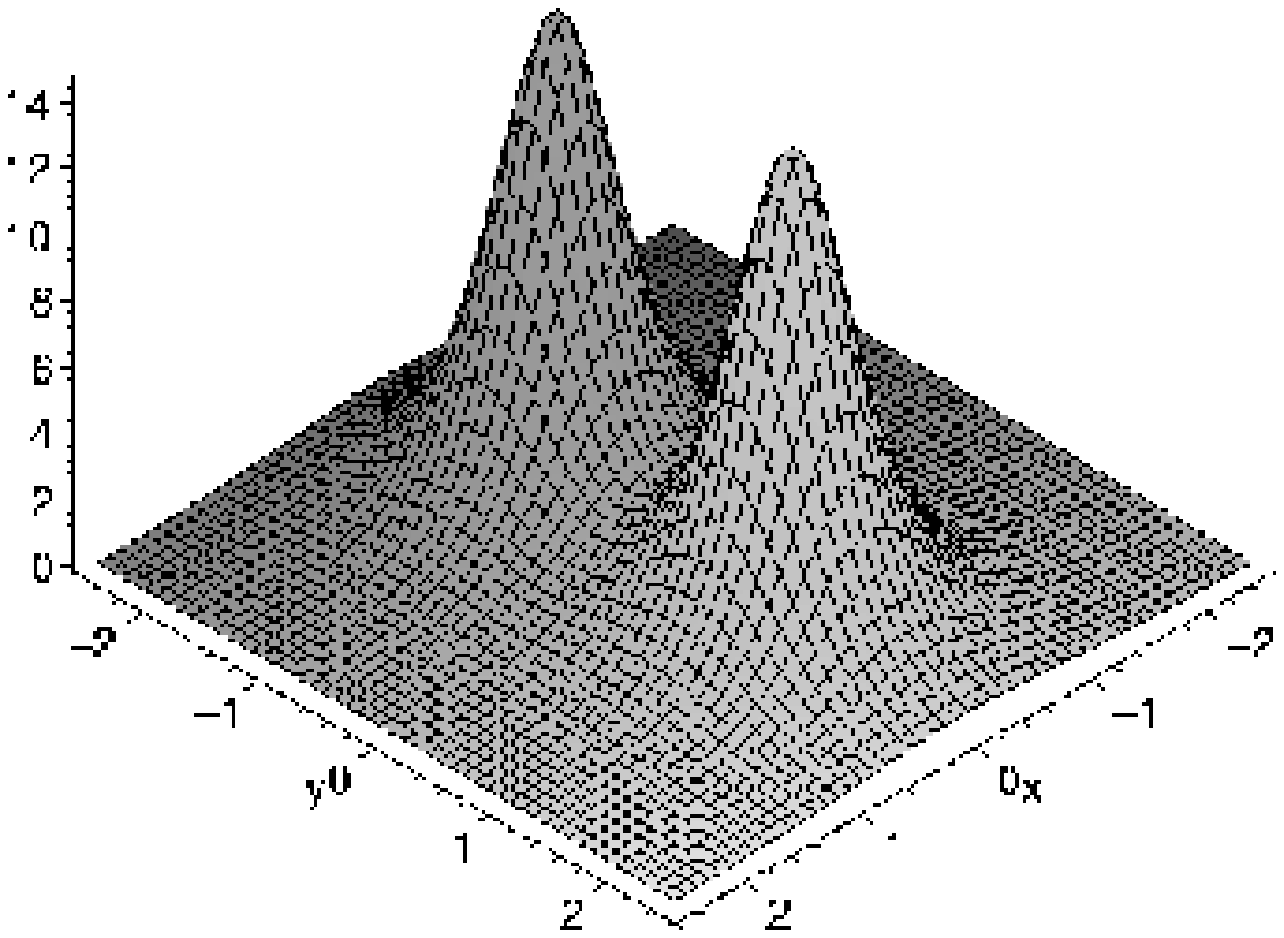}
\put(-245,100){$t=1$}
\vskip -.5cm
\caption{An elastic non-trivial scattering of a soliton-antisoliton
 configuration in flat spacetime.}
\label{fig-esa}
\end{figure}

\begin{figure}
\hskip .2cm
\epsfxsize=7cm\epsfysize=7cm\epsffile{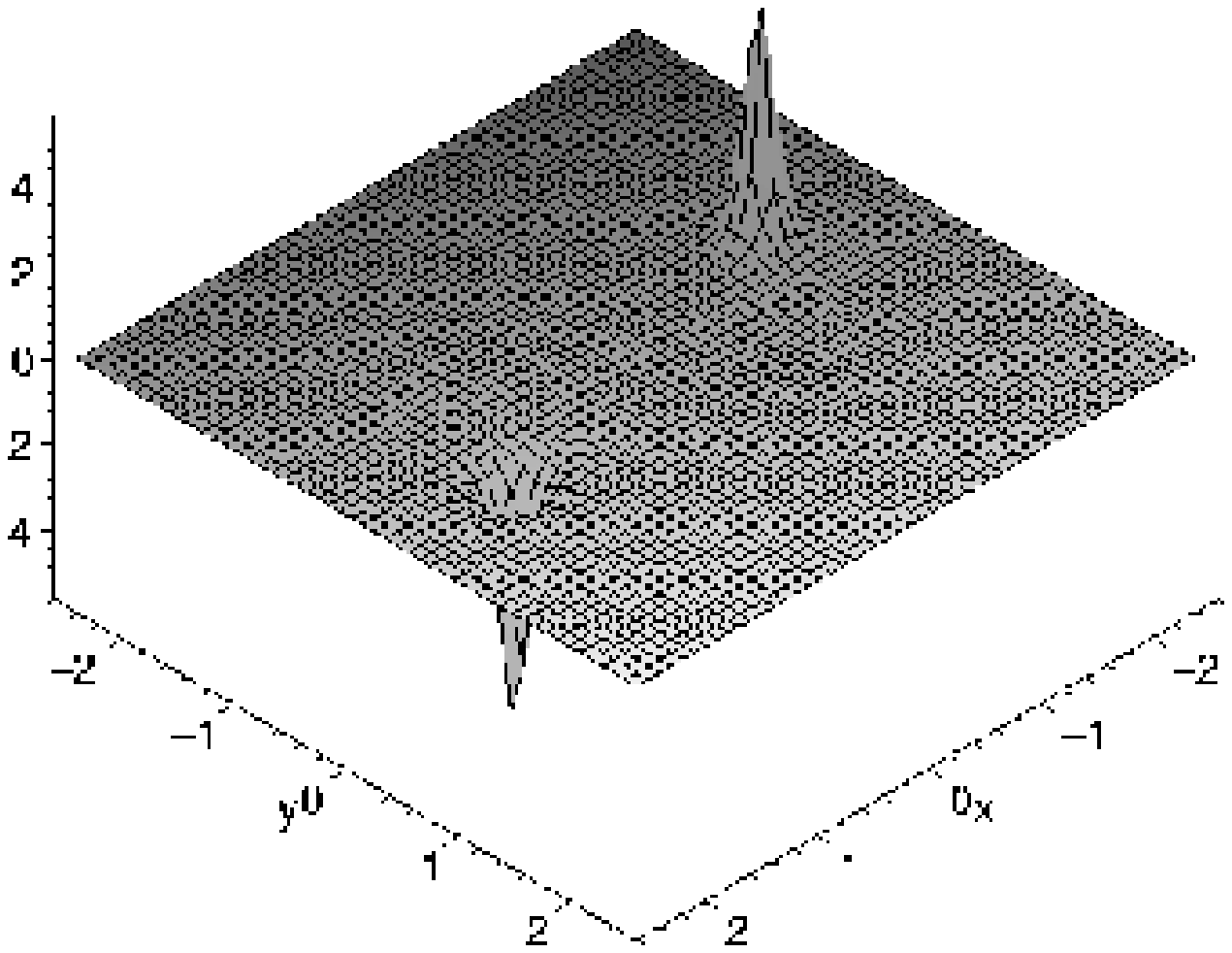}
\par
\vskip -1.9cm
\hskip 7cm
\epsfxsize=7cm\epsfysize=7cm\epsffile{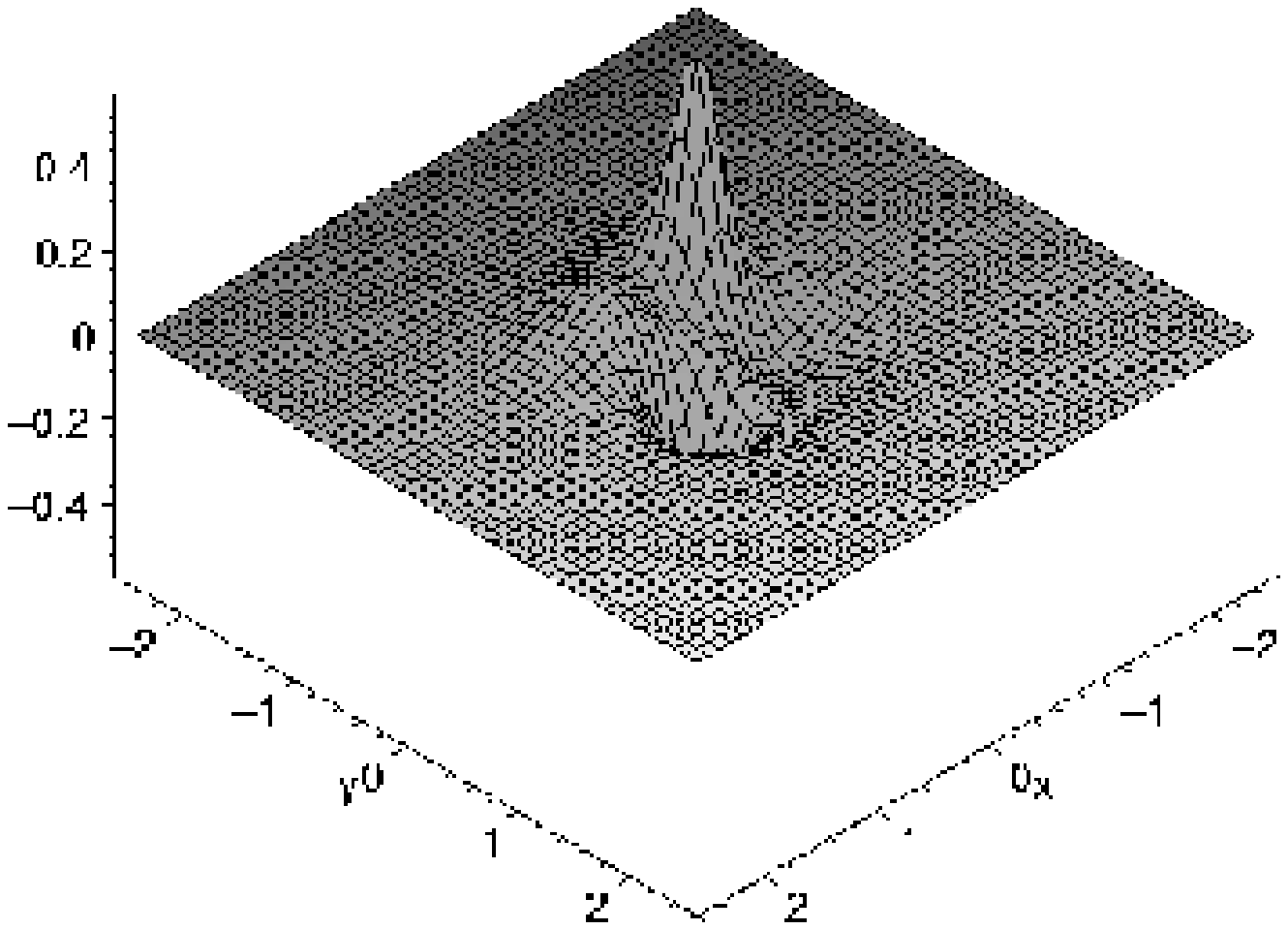}
\par
\hskip .2cm
\vskip -3cm
\epsfxsize=7cm\epsfysize=7cm\epsffile{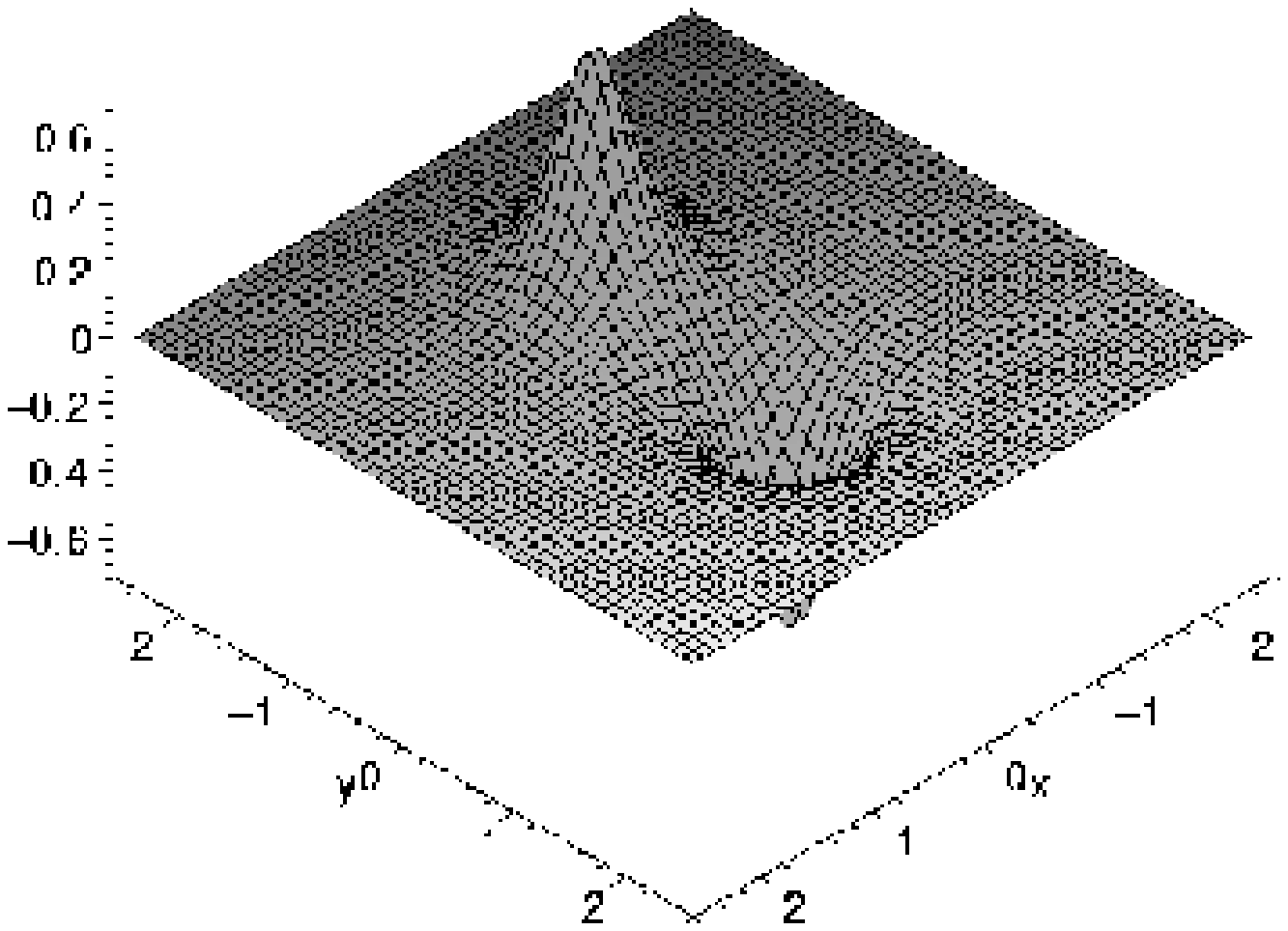}
\par
\vskip -1.8cm
\hskip 7cm
\epsfxsize=7cm\epsfysize=7cm\epsffile{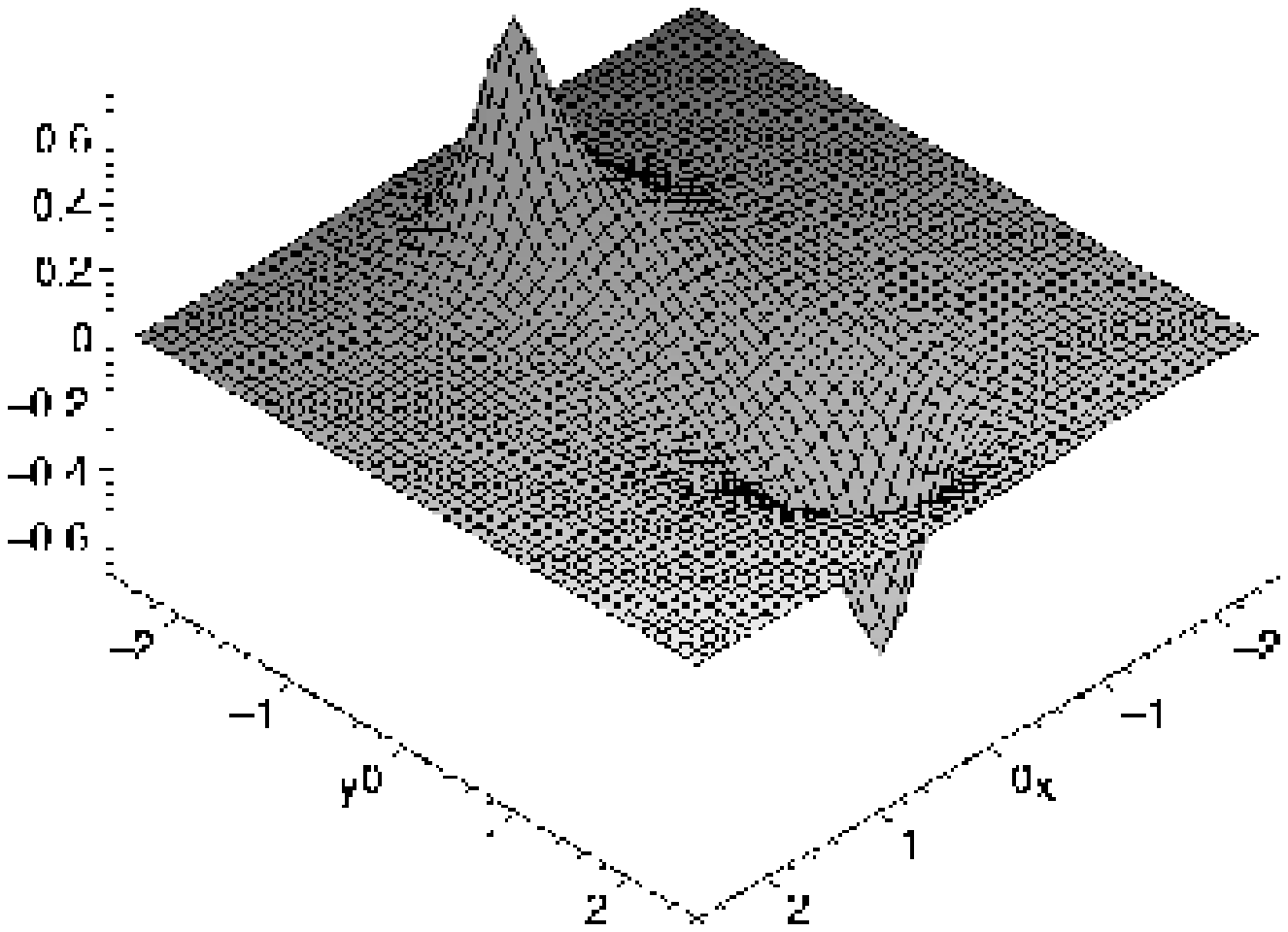}
\vskip -0.5cm \caption{Topological charge
density at increasing times for soliton-antisoliton elastic
scattering.} \label{fig-eto}
\end{figure}

\begin{itemize}
\item Lets take the simplest case where $f(z)=z$ and $h(\bar{z})=1$.
Roughly speaking the chiral field $J$ departs from its asymptotic
 value $J_0$  when $w \ra 0$ which implies that $z \ra \pm\sqrt{-2t}$.
More precisely, the two structures, for $t$ negative, are located
approximately at $x \approx \pm \sqrt{-2t}$ while, for $t$
positive, they are on the $y$-axis at $y \approx \pm \sqrt{2t}$.
Figure \ref{fig-esa} illustrates the scattering behavior of our
configuration
 near $t=0$.

The picture is consistent with the properties of the energy
density of the solution which is given by \be {\cal E}=16
\fr{2r^4+4r^2+4t^2(1+2r^2)-4t(x^2-y^2)+1}
{\left(2r^4+2r^2+4t(x^2-y^2)+4t^2+1\right)^2} \ee and is symmetric
under the interchange $t\ra -t, x\ra y, y \ra x$. The time
symmetry of the energy density confirms the lack of radiation.
However, the corresponding localized structures are not constant
of size; in fact, their height is proportional to $t^{-1}$ and
their radius is proportional to $\sqrt{t}$.

The projected topological charge ${\cal N'}$ is zero throughout the
 scattering process while the  topological density has an almost
identical distribution (up to a scale) to that of the energy
density as shown in Figure \ref{fig-eto}. Therefore, the
configuration represents a soliton and an antisoliton that are
clearly visible as distinct structures having respectively one and
minus one units of topological charge concentrated in a single
lump.

\end{itemize}

In the general case where $h=z^p$ and $h=\bar{z}^q$, the chiral field
$J$ departs from its asymptotic value $J_0$ when $w=z^{(p-1)}(2tp+z^N) \ra
0$ with $N=p+q+1$ which is true when either $z^{(p-1)}=0$ or $2tp+z^N=0$
and this is approximately where the lumps are located.
So $J$ represent a family of soliton-antisoliton solution which consists
of $(p-1)$ static soliton-like objects at the origin with $N$ others
accelerating toward them, scattering at an angle of $\pi/N$ and then
decelerating as they separate.

Note that in the non-integrable models (usually) there is an
attractive force between solitons of opposite topological charge
which implies that the initially well separated solitons and
antisolitons attract each other and eventually annihilate into a
wave of pure radiation which spreads with the velocity of light in
a direction perpendicular to the motion of the initial structures
as shown in \cite{Zakp}. However, it is known that the interaction
forces between solitons and antisolitons do depend on their
relative orientation in the internal space which implies that the
cross section for the soliton-antisoliton elastic scattering is
different than zero. In fact the proton-antiproton elastic
scattering is seen in a reasonable fraction of cases. Therefore
the analytic construction of families of soliton-antisoliton
configurations with elastic scattering properties obtained are of
immense importance since not only provide a major link between
integrable and non-integrable models but can actually
 have physical applications in the laboratories.

\section{ Conclusions}

The infinite number of conservation laws associated with a given
system place severe constraints upon possible soliton dynamics.
The construction of exact analytic multi-solitons with trivial
scattering properties is a result of such integrability
properties. In this paper soliton and soliton-antisoliton
structures have been presented for two planar integrable equations
related with the three-spatial dimensional self-dual
Yang-Mills-Higgs equations. These structures travel with
non-constant velocity, their size is non-constant and they
interact non-trivially. As we have already mentioned this
non-trivial scattering is not usual in an integrable theory but is
exceptional. Such results are useful for investigating the
connection between integrable and non-integrable systems which
possess multi-soliton solutions. In addition, they indicate the
likely occurrence of new phenomena in
 higher-dimensional soliton theory which are not present in
 one-dimensional systems.

The multi-pole ansatz have been extended to other planar
integrable systems in order  to obtain soliton solutions with
non-trivial dynamics. In particular,  a large class of solutions
of the Davey-Stewartson II equation has been constructed in
\cite{MS} which have an arbitrary rational localization in the
plane and describe typical interactions  consisting of head-on
collisions
 with a $\pi/2$ scattering angle.
Similarly, in \cite{VA} the discrete spectrum of the
non-stationary Schrodinger equation and  localized solutions of
the Kadomtsev-Petviashvili I
 equation  are studied via the inverse scattering transform.
It was shown that  there exist infinitely many real and rationally
decaying potentials  which correspond to a discrete spectrum whose
related eigenfunctions  have multiple poles in the spectral parameter;
while the resulting localized solutions of  the Kadomtsev-Petviashvili I
 behave as a collection  of individual humps with nonuniform dynamics.

It would be interesting to understand the role of higher poles
in algebraic-geometry approach like twistor theory (for example,
the function $\Psi$ given by (\ref{Psi-sca}) correspond to $n=2$ bundles),
 and also to investigate the construction of the corresponding solutions
and their dynamics  in de Sitter space. Finally, it would be
interesting to extend our construction in higher dimensional
gauged theories and investigate the scattering behavior of the
corresponding classical solutions and, also, consider and study
its noncommutative version (see, for example, Ref. \cite{LP}).

\section*{Acknowledgements}

Many thanks to the Royal Society and the National Hellenic
Research Foundation for a Study Visit grant.

\end{document}